\documentclass[a4paper,fleqn]{cas-sc}

\usepackage[numbers]{natbib}
\usepackage{booktabs}
\usepackage{caption} 
\usepackage{siunitx} 
\usepackage{array} 
\usepackage{tabularray}

\usepackage[caption=false]{subfig}
\usepackage{algorithm}
\usepackage{algpseudocode}

\hypersetup{
    colorlinks = true,
    citecolor = cyan,
    linkcolor = cyan,
    urlcolor = cyan,
    bookmarksopen = true,
    filecolor=cyan
}

\begin{document}
\let\WriteBookmarks\relax
\def\floatpagepagefraction{1}
\def\textpagefraction{.001}
\shorttitle{Learning to Route Electric Trucks Under Operational Uncertainty}
\shortauthors{Orfanoudakis S. et~al.}

\title [mode = title]{Learning to Route Electric Trucks Under Operational Uncertainty}

\author[1,2]{Stavros Orfanoudakis}[orcid=0000-0002-0767-9488]
\cormark[1]
\cortext[1]{Corresponding author email: s.orfanoudakis@tudelft.nl}
\credit{Conceptualization, Methodology, Software, Writing - Original Draft, Writing - Review \& Editing}

\affiliation[1]{organization={Delft University of Technology, Intelligent Electrical Power Grids},
            city={Delft},          
            country={The Netherlands}}

\affiliation[2]{organization={Massachusetts Institute of Technology, Center of Transportation and Logistics},
            city={Cambridge},
            state={MA},           
            country={United States of America}}
\author[2]{Ziyan Li}[orcid=0009-0009-5226-0430]
\credit{Conceptualization, Software, Writing - Review \& Editing}
\author[3]{Ruixiao Yang}[orcid]
\credit{Conceptualization, Writing - Original Draft, Writing - Review \& Editing}
\author[2]{Nikolay Aristov}[orcid=0009-0007-8832-1907]
\credit{Conceptualization, Software, Writing - Review \& Editing}
\author[1]{Pedro P. Vergara}[orcid=0000-0003-0852-0169]
\credit{Supervision, Funding acquisition, Writing - Review \& Editing}
\author[3]{Chuchu Fan}[orcid=0000-0003-4671-233X]
\credit{Supervision, Writing - Review \& Editing}
\author[2]{Elenna Dugundji}[orcid=0000-0001-9277-9485]
\credit{Conceptualization, Supervision, Writing - Review \& Editing}

\affiliation[3]{organization={Massachusetts Institute of Technology, Department of Aeronautics and Astronautics},
            city={Cambridge},
            state={MA},
            country={United States of America}}

\begin{abstract}
Electric truck operations require routing decisions that remain feasible under limited battery range, long charging times, travel and energy consumption, and competition for shared charging infrastructure. These features make electric truck routing a coupled logistics and energy problem, limiting the practicality of heuristics-based methods and rendering them computationally infeasible at scale. This paper proposes a learning-based framework for the stochastic electric truck routing under charging constraints and operational uncertainty. The problem, solved by Reinforcement Learning, is formulated as an event-driven semi-Markov decision process with shared charging resources, stochastic travel and energy requirements, and realistic nonlinear fast-charging behavior. To support learning in this setting, a graph-based representation of system state and feasible decisions is introduced, together with a rule-based action mask that restricts policies to operationally admissible actions; thus, improving training efficiency. Building on this formulation, an event-driven simulation environment is developed that supports both Reinforcement Learning and benchmarking against heuristic and mathematical programming baselines. Computational experiments across a range of fleet sizes show that the proposed learning-based algorithm consistently outperforms baselines and attains performance close to optimization benchmarks in many settings, while preserving high success rates under charging congestion and uncertainty.

\end{abstract}




\maketitle

\section{Introduction}

The electrification of transportation is transforming mobility and logistics systems by creating new opportunities to reduce emissions, improve energy efficiency, and lower dependence on fossil fuels~\cite{ZHANG2022103152}. Within freight transportation, electric trucks are particularly promising because they can reduce operating costs and local environmental impacts while supporting the decarbonization of regional and urban logistics. At the same time, their deployment changes the nature of routing and dispatching decisions~\cite{KUCUKOGLU2021107650}. Unlike conventional trucks, electric trucks operate under limited driving range, long charging times, and uneven charging infrastructure, making charging a central component of route planning rather than a secondary operational consideration. As a result, electric truck routing becomes a coupled logistics and energy management problem in which route feasibility and service performance depend not only on delivery sequences~\cite{osti_1615213}, but also on when, where, and how vehicles charge, especially in fleet settings with shared and capacity-constrained charging resources~\cite{BRAGIN2024104494}. 

This coupling becomes especially consequential in operational settings where routing decisions must be made dynamically and under uncertainty. In practice, electric truck operations are affected not only by travel distances and customer sequences, but also by charging timing, location, and duration~\cite {SPINELLI2026105480}. Public or shared charging stations may be capacity-constrained, subject to congestion, and used by multiple stakeholders, so a truck's feasible route depends in part on the behavior of other actors in the system. As a result, route feasibility and service performance are shaped jointly by battery state, charging availability, queueing delays, and stochastic travel and service conditions~\cite{KESKIN2021105060}. These features are amplified at the fleet level, where several trucks may compete for limited charging capacity~\cite{ev2gym} and where charging delays are endogenous rather than exogenously specified~\cite{WANG2025104932}. In such environments, methods for electric truck routing must do more than identify short paths, they must also support operationally feasible, time-sensitive decisions in systems where routing and charging are tightly interdependent.

The electric truck fleet routing problem (eTFRP)~\cite{ziyan2025} has evolved rapidly from a niche extension of the traditional vehicle routing problem (VRP) and electric VRP (eVRP) into a distinct research stream centered on the operational realities of electrified freight systems~\cite{cataldo_2023}. Optimization-based methods provide the main analytical foundation by embedding charging decisions, uncertainty, and energy-feasibility constraints directly into mathematical models~\cite{ziyan2025}. This line includes stochastic formulations for routing with simplified charging curves~\cite{SPINELLI2026105480}, robust models for heavy-duty electric trucks under uncertain energy consumption~\cite{AMIRI2023109108}, and conic mixed-integer programs that capture richer physical drivers of energy use such as road conditions, vehicle dynamics, payload, and distance~\cite{WANG2024123407}. These approaches offer strong interpretability and disciplined feasibility control, but their computational requirement grows rapidly with fleet size, uncertainty, and modeling realism, often requiring decomposition or approximation to remain tractable~\cite{lara_electric_2020}. Metaheuristics offer a more scalable alternative by combining charging-aware route construction, local improvement, and repair mechanisms within flexible search procedures. Variable-neighborhood search~\cite{euchi_hybrid_2023}, memetic~\cite{dong_dynamic_2023}, and ant-colony-based methods~\cite{9409782} have shown that large instances can be handled more efficiently in practice, although their performance often depends on instance structure and parameter tuning and is harder to characterize systematically under uncertainty.

In parallel, Reinforcement Learning (RL) has emerged as a promising direction~\cite{evgnn} because it enables search across instances during training and enables rapid inference at deployment time~\citep{kool2018attention}. For classical routing problems, attention-based and transformer-style policies, together with improved rollout and symmetry-based training strategies, have demonstrated that learned policies can achieve strong performance without hand-crafted search rules~\citep{yang2025neural,Sym-NCO}. Recent studies have extended these ideas to charging-aware, stochastic, and safety-aware electric vehicle routing settings~\cite{BASSO2022102496}. However, much of this literature still remains closer to simple eVRP settings than to more realistic fleet-level electric truck operations with shared charging resources, traffic congestion, and truck-specific operational constraints~\citep{lin2025, TANG2023121711}. In addition, learning-based approaches rarely embed explicit feasibility control in the decision process, and comparisons across optimization, heuristic, and learning methods are often difficult to interpret because they rely on different assumptions about uncertainty, charging behavior, and temporal dynamics~\cite{lin2022}. These gaps motivate the need for a truck-oriented, learning-compatible framework that combines a realistic operational structure, feasibility-aware decision-making, and systematic benchmarking within a common evaluation pipeline.

These gaps point to a clear need for methods that retain the operational realism of electric truck routing while remaining scalable enough for repeated, time-sensitive decision making. In response, this paper develops a truck-oriented and learning-compatible framework\footnote{The project repository is available at \url{https://github.com/StavrosOrf/eTruckRouting}.} for the eTFRP under charging constraints and uncertainty. The problem is formulated to explicitly capture fleet-level interactions through shared charging infrastructure, stochastic travel and energy conditions, and realistic charging behavior, while preserving the structure needed for sequential decision-making. On this basis, Graph Proximal Policy Optimization (GraphPPO) is introduced as a graph neural network (GNN)-based~\cite{kipf} RL approach that combines a structured state--action representation with an explicit state-based feasibility action mask, allowing routing and charging decisions to be learned without discarding the operational constraints that define the problem. To assess both the strengths and the limitations of GraphPPO, a common simulation and benchmarking environment is further developed for direct comparison with heuristic and optimization-based methods under consistent assumptions. Across the main benchmark, the proposed method is shown to remain consistently close to the mathematical optimization benchmark over a broad range of fleet sizes, to outperform the other learning-based baselines in both solution quality and robustness, and to become particularly competitive in the largest and most congested settings. In the single-truck eVRP setting, near-optimal performance is also retained while requiring substantially less online computation than mathematical optimization.

More specifically, the paper's contributions can be summarized as follows:
\begin{itemize}
    \item The eVRP and eTFRP are formulated as event-driven semi-Markov decision processes with shared charging resources, stochastic travel and energy consumption, and realistic nonlinear charging behavior, thereby moving beyond simplified abstractions toward a truck-oriented operational setting compatible with sequential decision making under uncertainty.

    \item A graph-based state and action representation is introduced together with a state-based action feasibility mask, so that interactions among trucks, deliveries, and charging infrastructure are encoded explicitly while decisions remain restricted to operationally admissible actions.

    \item GraphPPO is proposed as a graph-based actor--critic RL method designed for routing and charging under fleet-level coupling, uncertainty, and variable action spaces. Across the main eTFRP benchmark, near-benchmark performance is maintained as problem size increases, while stronger reliability and scalability are obtained than with the generic PPO and MaskPPO baselines.

    \item A complete event-driven simulation and benchmarking environment is developed using real transportation networks from California, together with an extensive computational study that evaluates scalability, robustness, and transferability relative to heuristic and mathematical programming baselines. The results show strong competitiveness in large congested fleet settings, near-optimal quality in the single-vehicle eVRP case, and meaningful zero-shot generalization across unseen problem configurations.
\end{itemize}

\section{Related Work}

The eTFRP introduces routing challenges beyond the classical VRP. In addition to determining customer visit sequences, feasible plans must account for when and where vehicles recharge while remaining robust to uncertainty in travel times and energy consumption. These interactions are particularly pronounced for battery-electric trucks, whose operating costs and route feasibility are more sensitive to payload, road conditions, and charging availability. Accordingly, the literature on electric vehicle and electric truck routing has evolved along three main directions: optimization-based approaches that explicitly model charging and uncertainty, metaheuristics that emphasize scalability and flexibility, and learning-based methods that seek fast, adaptive decision-making. Table~\ref{tab:taxonomy} summarizes representative studies across these directions. 

\begin{table*}
\scriptsize
\centering
\setlength{\tabcolsep}{4pt}
\renewcommand{\arraystretch}{1.15}
\caption{Overview of representative studies on eVRP and eTFRP. The table compares prior work by problem variant, modeled uncertainty, method family, and key operational realism. ``---'' indicates that the corresponding aspect was not considered in the study. TW refers to the time-window variants of the VRP problem, and TSP to the traveling salesman problem.}
\label{tab:taxonomy}
\begin{tabular}{p{0.5cm}p{1.8cm}p{3.0cm}p{2.5cm}p{7cm}}
\toprule
\textbf{Ref.} &
\textbf{Problem variant} &
\textbf{Uncertain Variables} &
\textbf{Method family} &
\textbf{Key modeling realism} \\
\midrule

\cite{BRAGIN2024104494} &
eTFRP &
--- &
MILP &
Surrogate level-based Lagrangian relaxation \\

\cite{SPINELLI2026105480} &
eTFRP &
Energy consumption &
Iterated Local Search &
Travel time + recharging operations \\

\cite{WANG2025104932} &
eTFRP-TW &
--- &
Iterated Local Search &
Time windows + charging queues + nonlinear ch. curves \\

\cite{cataldo_2023} &
eTFRP-TW &
--- &
MILP &
Time windows + nonlinear charging curves + battery degradation \\

\cite{ziyan2025} &
eTFRP &
Charger queue &
SDDP &
Energy consumption, asymmetrical graph \\

\cite{WANG2024123407} &
eVRP &
Energy consumption &
MIP &
Energy consumption, asymmetrical graph \\

\midrule


\cite{AMIRI2023109108} &
eTFRP-TW &
Energy consumption &
Metaheuristic &
Robustness to consumption variability \\

\cite{euchi_hybrid_2023} &
eTFRP &
--- &
Metaheuristic (VNS) &
Cost/distance objectives; fleet sizing insight \\

\cite{dong_dynamic_2023} &
Dynamic eTFRP &
Demand arrivals &
Metaheuristic (Mem.) &
Mid-route recharging + new requests \\

\cite{9409782} &
eTFRP &
--- &
Metaheuristic (ACO) &
Bilevel: routing + charging repair \\

\cite{https://doi.org/10.1155/2021/6635749} &
Dynamic eVRP &
--- &
Metaheuristic (Mem.) &
Adaptive local search + random immigrants \\

\cite{https://doi.org/10.1049/pel2.12555} &
eVRP &
--- &
Metaheuristic &
Battery + charging schedule + service constraints \\

\midrule
\cite{kool2018attention} &
TSP/VRP &
--- &
RL&
Attention-based policy \\

\cite{BASSO2022102496} &
Dynamic-eVRP &
Energy consumption &
Safe RL &
Risk-aware routing; offline Monte Carlo \\

\cite{lin2025} &
eVRP&
--- &
RL (Q-Learning) &
Time aspect noted \\

\cite{TANG2023121711} &
eVRP &
--- &
RL (Q-Learning)&
Dynamics + road + charging losses \\

\cite{lin2022} &
eVRP-TW &
--- &
RL + heuristic &
Q-learning selects heuristics; charging repair \\

\cite{NEURIPS2018_9fb4651c} &
VRP &
--- &
RL (policy gradient) &
Reward-driven learning + feasibility rules \\

\cite{NEURIPS2023_9bae70d3} &
TSP &
--- &
RL + Local search &
Learned $k$-opt; guided exploration \\

\cite{10608117} &
eVRP-TW &
--- &
RL (REINFORCE) &
Autoregressive encoder--decoder \\

\cite{11016767} &
eVRP &
--- &
RL (attention) &
Heterogeneous attention mechanism \\

\cite{10310266} &
eVRP &
--- &
Multi-agent RL &
Agent interactions; grid/power-network aspect \\

\cite{ALQAHTANI2022122626} &
eVRP &
Energy consumption &
RL&
Dispatch across locations; energy-delivery framing \\

\midrule
\textbf{Ours} &
eVRP \& eTFRP &
Energy consumption, traffic, and charger queue &
RL (PPO) + GNN &
Charging curve, traffic congestion, asymmetric transportation graph,  \\

\bottomrule
\end{tabular}
\end{table*}

\subsection{Classic Optimization Approaches}

Optimization-based approaches form the analytical backbone of electric truck routing by embedding charging decisions, uncertainty, and energy-feasibility constraints directly into mathematical formulations. A common objective in this direction is to preserve operational realism by explicitly coupling routing and charging, often at the expense of computational tractability. Stochastic formulations illustrate this direction by separating route design from charging recourse under uncertainty. For example, \cite{SPINELLI2026105480} models customer sequencing in a first stage and charging decisions in a second stage to minimize expected route duration. Robust optimization provides a complementary perspective for freight applications, where battery depletion is operationally unacceptable, as shown by \cite{AMIRI2023109108} for heavy-duty electric trucks under uncertain energy consumption. Conic and mixed-integer formulations further increase modeling fidelity by incorporating realistic energy-consumption drivers such as road conditions, vehicle dynamics, payload, and distance~\citep{WANG2024123407}. Although these models offer strong interpretability and disciplined feasibility control, their execution time grows rapidly with fleet size, uncertainty, and physical realism, motivating decomposition~\citep{ziyan2025} and approximation~\cite{lara_electric_2020} techniques to improve scalability in multistage settings using Stochastic Dual Dynamic Programming (SDDP). 
Optimization-based methods provide a principled foundation for eTFRP, but their computational cost can become prohibitive when large fleets, detailed charging behavior, and uncertainty must be modeled jointly. In contrast, our approach preserves these operational interactions in a learning-compatible formulation that supports faster decision-making and direct comparison with optimization baselines within a common simulation setting.

\subsection{Metaheuristic and Population-Based Approaches}

To reduce the computational burden of high-fidelity stochastic and robust formulations, many studies adopt metaheuristics that scale better to larger problem instances while accommodating charging-related constraints. A common strategy is to integrate routing and charging decisions within flexible search procedures, including methods that jointly optimize trip time, cost, charging schedules, and service constraints \citep{https://doi.org/10.1049/pel2.12555}. Variable-neighborhood search (VNS) and other hybrid improvement heuristics have also been used to address capacitated electric truck routing while targeting cost and distance objectives \citep{euchi_hybrid_2023}. When dynamic conditions are present, evolutionary and memetic methods become especially attractive because they can combine local search with adaptive diversification mechanisms to handle mid-route recharging and newly arriving requests \citep{dong_dynamic_2023,https://doi.org/10.1155/2021/6635749}. Bilevel ant-colony optimization (ACO) approaches provide another perspective by separating customer sequencing from charging-feasibility repair \citep{9409782}. 
Overall, metaheuristics offer a practical balance between flexibility and computational speed, but their performance is often instance- and tuning-dependent, and their treatment of uncertainty and feasibility is typically embedded in problem-specific repair logic. In contrast, our method encodes operational feasibility directly into the decision process via a rule-based action mask and systematically evaluates performance in multiple stochastic settings of varying size.

\subsection{Reinforcement Learning Approaches}

Here, the research progress on RL-based routing is split into advances in classic routing (VRP and TSP) and its electric vehicle extension (eVRP).

\subsubsection{RL for classical routing problems}

Recent progress in RL for routing has been driven by learned heuristics that enable combinatorial search across instances and fast decision-making at deployment time, both of which are essential for freight operations that require frequent replanning. A central paradigm is the constructive encoder--decoder policy, in which routes are built sequentially using attention mechanisms. The attention model of \cite{kool2018attention} demonstrated that such policies can achieve strong performance with policy-gradient training and rollout-based baselines, while related work showed that a single policy can learn near-optimal routing decisions over broad instance distributions using only reward signals and feasibility rules \citep{NEURIPS2018_9fb4651c}. Subsequent research has focused on improving training stability and generalization through diverse rollouts and symmetry-aware regularization, leveraging the invariances of routing problems to improve robustness beyond the training distribution \citep{Sym-NCO}. Beyond constructive policies, learning-to-search approaches have sought to emulate classical improvement heuristics by learning local search moves directly; for example, \cite{NEURIPS2023_9bae70d3} learns flexible transformations and uses guided exploration to navigate complex search spaces. More broadly, these studies show that RL can serve as an effective framework for routing, particularly when fast inference and repeated decision-making are required.

\subsubsection{RL for electric vehicle routing}

Building on advances in learning-based routing, recent studies have extended RL to settings where routing decisions must remain energy-feasible and explicitly account for charging actions. One stream adapts constructive attention-based policies to eVRP variants, such as eVRP-TW (with Time-Windows), demonstrating that end-to-end neural policies can handle charging and time-window constraints with competitive performance on benchmark instances \citep{lin2025,10608117}. A second stream increases modeling realism by incorporating vehicle dynamics, road conditions, and charging losses into the learning process through transformer-based or attention-enhanced architectures tailored to eVRP settings \citep{TANG2023121711,11016767}. A third stream focuses on uncertainty and safety, which are especially important in electric freight because battery depletion carries substantial operational risk; for example, safe RL formulations have been proposed for dynamic stochastic settings with uncertain demand and energy consumption \citep{BASSO2022102496}. Hybrid and multi-agent approaches further broaden the design space by combining RL with heuristic selection, charging-repair mechanisms, or explicit modeling of agent interactions and grid-related considerations \citep{lin2022,10310266,ALQAHTANI2022122626}. 

Overall, RL-based electric vehicle routing research has developed rapidly, but most existing methods address only part of the problem addressed here. Prior studies typically focus on single-vehicle or unrealistic eVRP settings, simplified charging assumptions, or isolated aspects such as safety, heuristic selection, or multi-agent coordination. By contrast, the proposed method is designed for fleet-level electric truck routing with shared charging resources, combines explicit feasibility-aware action selection with a graph-based learning architecture, and evaluates these decisions under uncertainty within a common benchmarking environment that also includes heuristic and optimization baselines.

\section{The Electric Truck Fleet Routing and Charging Problem}\label{sec:problem}

In this section, the eTFRP with shared charging resources and stochastic travel times and energy consumption is formalized. An event-driven simulation platform that models the eTFRP is also described, enabling detailed system-level evaluation and comparison of different solution approaches.

\subsection{Mathematical Formulation}\label{sec:problem_definition}

The eTFRP consists of a directed road network graph $G=(\mathcal{V},\mathcal{E})$ constructed from synthetic or real transportation data, where the node set $\mathcal{V}$ contains the depot, delivery, and charging stations locations, denoted by $\mathcal{D}\subseteq\mathcal{V}$ and $\mathcal{C}\subseteq\mathcal{V}$, respectively.
Each directed edge $(u,v)\in\mathcal{E}$ is associated with a deterministic base travel time $\tau_{uv}$ (hours) and a deterministic base energy requirement $e_{uv}$ (kWh), obtained from pre-computed shortest-path distances between each node.
These base quantities define the nominal road network cost structure used throughout the formulation.
Note that the transportation graph is not assumed to be symmetric. Consequently, traversing the edge $(u,v)$ may incur different travel time and energy consumption than traversing $(v,u)$, capturing direction-dependent effects such as road gradients, geometry, and traffic patterns~\cite{ziyan2025}.

Each charging station $c\in\mathcal{C}$ is characterized by its peak charging power $\overline{p}_c$ (kW), charging efficiency $\eta_c\in(0,1]$, and an integer port capacity $\kappa_c$, which limits the number of trucks that may charge simultaneously at that station. This capacity parameter is essential in fleet settings because it creates resource competition when multiple trucks arrive to charge at the same station during overlapping time intervals.

\begin{figure} 
\centering
     \subfloat[Electric Vehicle Routing Problem]{
         \centering
         \includegraphics[width=0.4\linewidth]{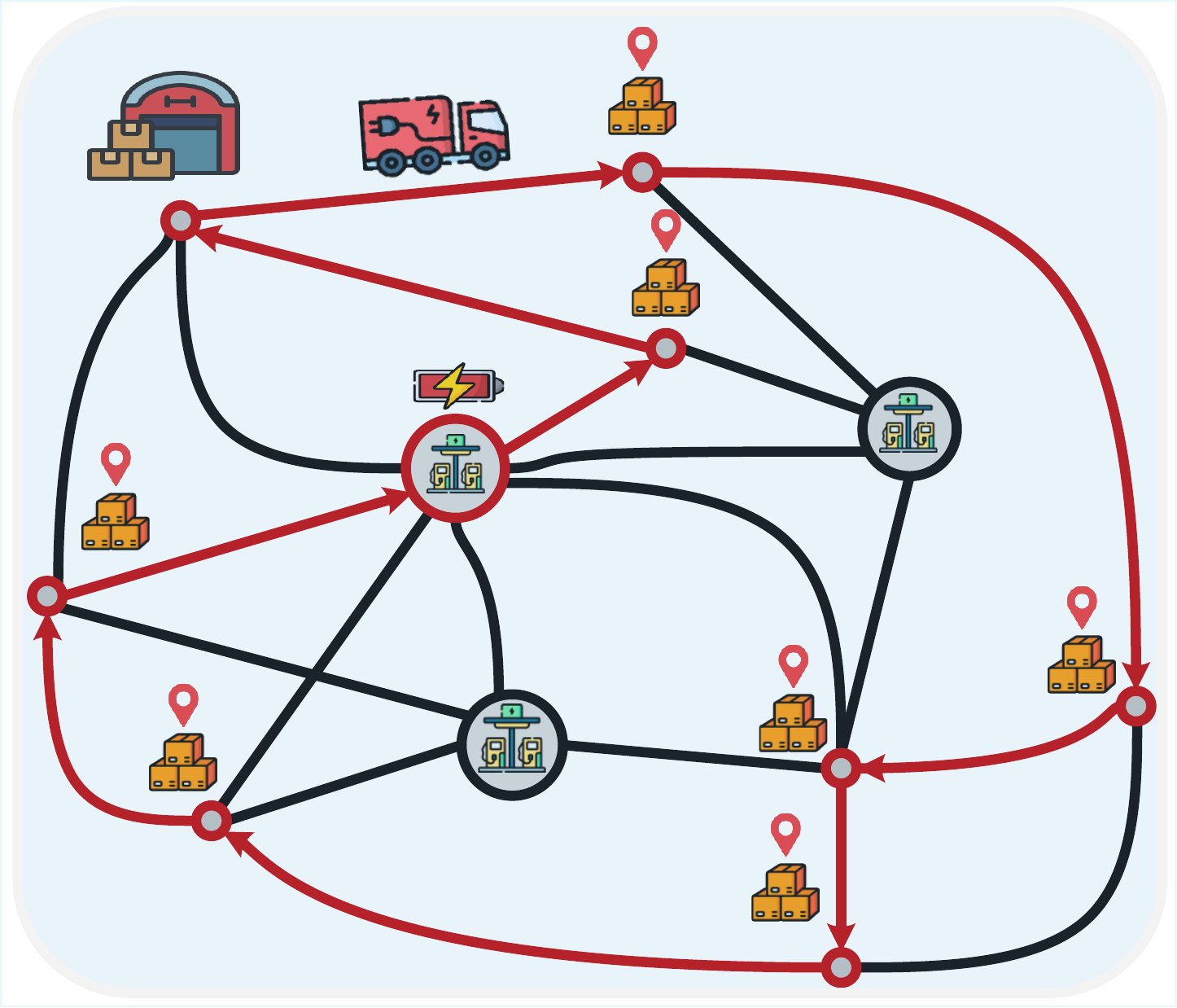}
         \label{fig:ov_a}
         }
     \subfloat[Electric Truck Fleet Routing Problem]{
         \centering
         \includegraphics[width=0.4\linewidth]{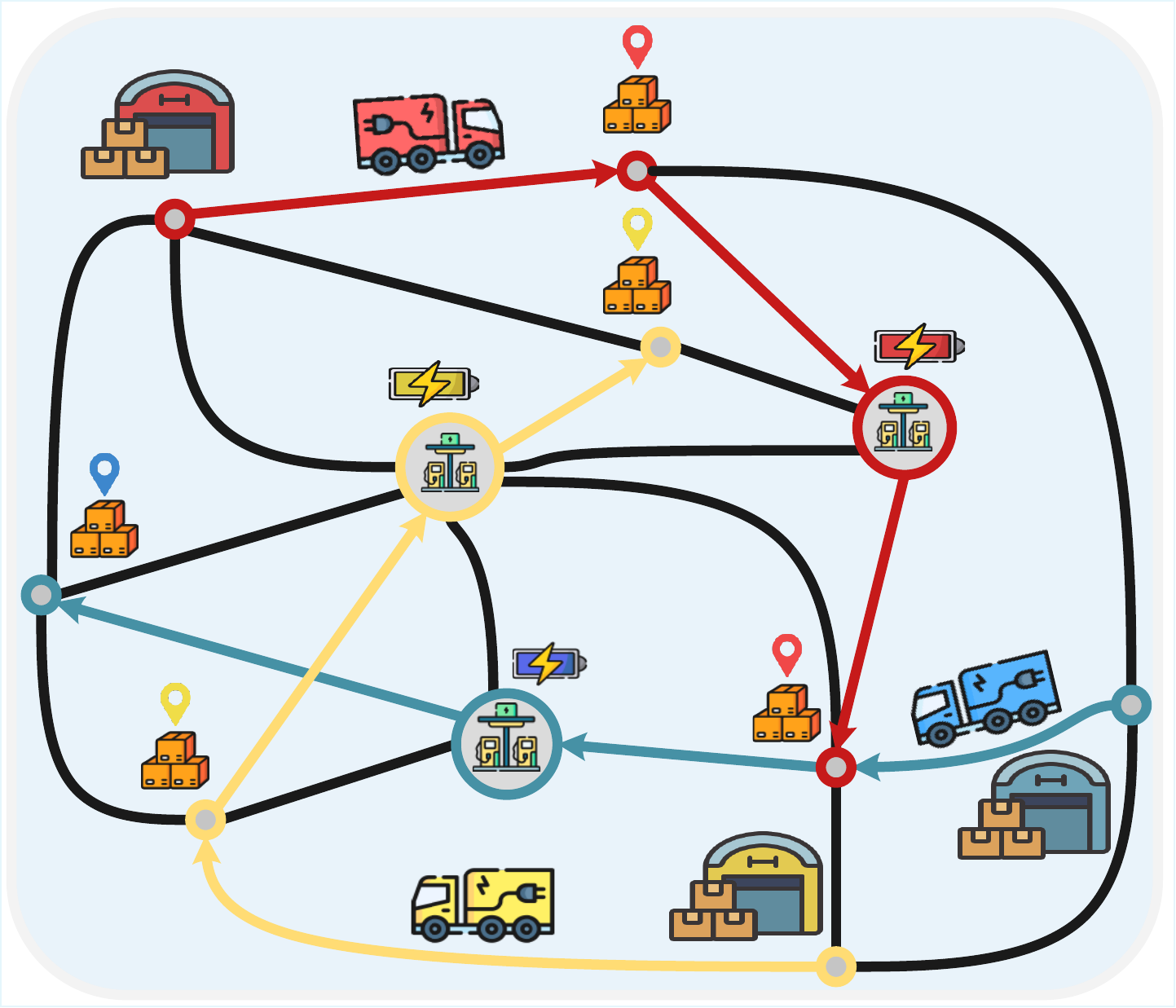}
         \label{fig:ov_b}
         }
        \caption{Comparison of the classic eVRP and the eTFRP with shared charging resources. (a) In eVRP, routing is typically planned for a single vehicle (or independently across vehicles), where charging stops are inserted to maintain energy feasibility along a prescribed tour. (b) In eTFRP, multiple trucks operate concurrently and compete for limited charging capacity, so route execution is coupled through shared stations and potential queuing delays.
        }
        \label{fig:overview}
\end{figure}

The fleet consists of $I$ electric trucks indexed by $\mathcal{I}=\{1,\ldots,I\}$. Each truck $i \in \mathcal{I}$ has battery capacity $\overline{b}_i$ (kWh) and must maintain a minimum energy $\underline{b}_i$ (kWh) to avoid fully discharging and preserve operational safety margins. The initial battery energy of truck $i$ is denoted by $b_i^0$, with the starting value set either to full capacity or sampled from a distribution, depending on the experimental configuration. Each truck starts at a designated depot (or start) node $d_{i}^0\in\mathcal{V}$, which determines the route's initial location. It is possible that every truck starts from a different depot, mirroring the operation of real freight distribution networks~\cite{MOGALE2025111315}.

Delivery points are specified as pre-assigned sets per truck. 
For each truck $i\in\mathcal{I}$, the assigned delivery set is denoted by $\mathcal{D}_i=\{d_{i}^1,\ldots,d_{i}^{K_i}\}$, where $K_i$ is the number of deliveries assigned to truck $i$. In the fixed delivery order mode (sequential), each truck must serve these deliveries in the given order, whereas in the flexible mode (similar to classic eVRP), each truck may choose the order of the remaining deliveries in $\mathcal{D}_i$ to minimize travel time.
Delivery sequences are validated to ensure that each path is feasible under conservative energy assumptions, which may require one or more intermediate charging stops. 

The fixed delivery order in the eTFRP problem differs from that in standard single-vehicle eVRP formulations in two structurally important ways.
First, deliveries are pre-assigned at the truck level (with either sequential or flexible ordering), shifting the operational burden toward energy-feasible execution rather than delivery-to-vehicle assignment. Additionally, in eTFRP, trucks are not necessarily required to return to the initial depot.
Second, charging is a shared resource with finite simultaneous service capacity, so that charging delays depend on the joint behavior of the fleet and cannot be represented as exogenous constants.
Figure~\ref{fig:overview} illustrates this distinction by contrasting the single-vehicle eVRP abstraction with a fleet-level setting in which interactions arise through shared charging infrastructure and associated congestion effects. 

\subsubsection{Sources of Uncertainty}
\label{sec:unc}

Uncertainty enters the problem through exogenous stochasticity in travel times, energy consumption, and unloading durations~\cite{lombard2018modelling}. Travel time on an edge $(u,v)$ departing at time $t$ is modeled as a truncated Gaussian~\cite{RODRIGUES2018636},
\begin{equation}
\tilde{\tau}_{uv}(t) \sim \mathcal{N}\!\big(\tau_{uv},\sigma_{uv}^2(t)\big),
\label{eq:unc_travel}
\end{equation}
where $\sigma_{uv}(t)$ increases during rush-hour periods based on the fraction of the journey interval overlapping rush windows, e.g., in the morning and evening.
Stochastic energy consumption is correlated with traffic to reflect stop-and-go inefficiencies or other truck battery-related nonlinearities, using $\tilde{b}_{uv} = b_{uv}\,\xi_{uv}$. The coefficient $\xi_{uv}$ is defined as:
\begin{equation}
\xi_{uv} =
\operatorname{clip}\!\Big(
1 + 0.5(\tilde{\tau}_{uv}/\tau_{uv}-1) + \varepsilon,\ \underline{\xi},\ \overline{\xi}
\Big),
\label{eq:unc_energy}
\end{equation}
with $\varepsilon$ representing additional noise and clipping bounds $[\underline\xi,\overline\xi]$ that prevent unrealistic extreme traveling time values. 
Unloading times at delivery nodes can also be stochastic (with realized time $\tilde{q}$), reflecting longer delays during business hours due to facility congestion and dock availability. In this setting, delivery points are assumed to be operational throughout the whole day.

Another source of uncertainty is the endogenous congestion at charging stations. Even if exogenous variables were fixed, the waiting time experienced at a station depends on the arrival times and charging durations of other trucks, making queuing delays a coupled, fleet-level problem. In eTFRP, stations maintain first-come, first-served waitlists and admit trucks to charging ports only when capacity becomes available.

\subsubsection{Objective function and constraints}\label{sec:obj_constraints}

eTFRP is a sequential decision-making problem that requires real-time decisions.
At each decision step $t \in \mathcal{T}$, the operator selects the next action for the active truck, which can be navigating to a delivery location (the next stop in sequential delivery mode or any remaining stop in flexible mode), navigating to a charging station, or, if the truck is currently at a station with access to a charging port, selecting a charging duration $h\in\mathcal{H}$.

Let $\tilde{\tau}^{(i)}_{k}$ denote the realized travel time for the $k$-th travel segment of truck $i$, let $h^{(i)}_{k}$ denote the duration of its $k$-th charging session, let $w^{(i)}_{k}$ denote the waiting time incurred before that charging session due to the congestion at the selected station, and let $\tilde{q}^{(i)}_{k}$ denote the realized unloading time for its $k$-th served delivery. 
The objective is to design a policy $\pi$ that minimizes the expected total time required for the whole fleet to complete all assigned deliveries, formally defined as:
\begin{align}
\min_{\pi}\quad
& \mathbb{E}\!\left[\sum_{i\in\mathcal{I}}
\left(
\sum_{k}\tilde{\tau}^{(i)}_{k}
+
\sum_{k} h^{(i)}_{k}
+
\sum_{k} w^{(i)}_{k}
+
\sum_{k}\tilde{q}^{(i)}_{k}
\right)\right]
\label{eq:obj_time}
\\
\text{s.t.:} \quad
& \underline{b}_i \le b_i(t) \le \overline{b}_i && \forall i\in\mathcal{I}, \ \forall t \in \mathcal{T},
\label{eq:con_battery_bounds}
\\
& \alpha\cdot e_{n_i(t),v} < b_i(t) && \forall i\in\mathcal{I},\ \forall t \in \mathcal{T},
| \; i\mapsto \text{navigates}(v),
\label{eq:con_energy_headroom}
\\
& b_i\!\left(t+\tilde{\tau}_{n_i(t),v}\right)= b_i(t)-\tilde{e}_{n_i(t),v} && \forall i\in\mathcal{I},\ \forall t \in \mathcal{T}\ | \; i\mapsto \text{navigates}(v),
\label{eq:con_energy_update_travel}
\\
& b_i(t+h_k^{(i)})=\min\!\Big\{\overline{b}_i,\ b_i(t)+ \int_{t}^{t+h}\eta_c \cdot P_c\!\left(\frac{b_i(t)}{\overline{b}_i}\right)\, dt,\Big\} && \forall i\in\mathcal{I},\ \forall t \in \mathcal{T}\ | \; i\mapsto \text{charges}(c,h),
\label{eq:con_energy_update_charge}
\\
& |\mathcal{M}_c(t)| \le \kappa_c && \forall c\in\mathcal{C}, \ \forall t \in \mathcal{T}.
\label{eq:con_ports}
\end{align}
Here $b_i(t)$ denotes the battery energy of truck $i$ at time $t$, $n_i(t)\in\mathcal{V}$ denotes its current location node, and $\mathcal{M}_c(t)\subseteq\mathcal{I}$ denotes the set of trucks occupying charging ports at station $c$ at time $t$. 
Constraints~\eqref{eq:con_battery_bounds} and \eqref{eq:con_energy_headroom} enforce energy feasibility throughout the operational day. The battery bound in~\eqref{eq:con_battery_bounds} maintains a minimum buffer $\underline b_i$ and prevents exceeding capacity $\overline{b}_i$. In detail, constraint~\eqref{eq:con_energy_headroom} requires that any selected navigation action to a node $u$ is feasible under a safety multiplier $\alpha\ge 1$, which conservatively accounts for energy uncertainty. 
The battery evolution is captured explicitly by~\eqref{eq:con_energy_update_travel} and \eqref{eq:con_energy_update_charge}.
In~\eqref{eq:con_energy_update_travel}, $\tilde{e}_{uv}$ and $\tilde{\tau}_{uv}$ denote the realized energy consumption and travel time on edge $(u,v)$. In~\eqref{eq:con_energy_update_charge}, the function $P_c(\cdot)$ returns the energy added over duration $h$ at station $c$ given charging efficiency $\eta_c$ and the station-specific max charging power $\overline{p}_c$. Its explicit charging curve equation is specified in Section~\ref{sec:charging_model}.

Constraint~\eqref{eq:con_ports} captures the coupling introduced by shared charging resources. In particular, \eqref{eq:con_ports} limits the number of trucks that can charge simultaneously at station $c$ and induces a first-come, first-served waiting time $w_k^{(i)}$ when a truck arrives while all $\kappa_c$ ports are occupied. 
By doing that, the realized waiting terms $\{w_k^{(i)}\}$ in~\eqref{eq:obj_time} are determined by the interaction of simultaneous routing and charging decisions across the fleet.

\subsection{Event-driven Simulation Environment}\label{sec:sim_env}

To enable reproducible evaluation of different eTFRP solutions, the mathematical problem is transformed into an event-driven simulation environment that executes fleet operations in continuous time. 
The simulator maintains a priority queue of timestamped events ($t\in \mathcal{T}$), and advances time by repeatedly processing the earliest event in the queue. This design avoids fixed-step discretization artifacts and naturally supports heterogeneous action durations, including travel segments of different lengths, charging sessions of variable duration, and unloading times.
At each event, the simulator updates the system state (truck locations, battery levels, station occupancies, and queues) and schedules subsequent events induced by the executed action, such as arrival at the next node, completion of charging, or completion of unloading.
The simulator environment uses the Gymnasium API~\cite{brockman2016openai} and is suitable for the development and testing of RL algorithms while also supporting mathematical programming and heuristic planning algorithms.

\begin{figure}
    \centering
    \includegraphics[width=0.8\linewidth]{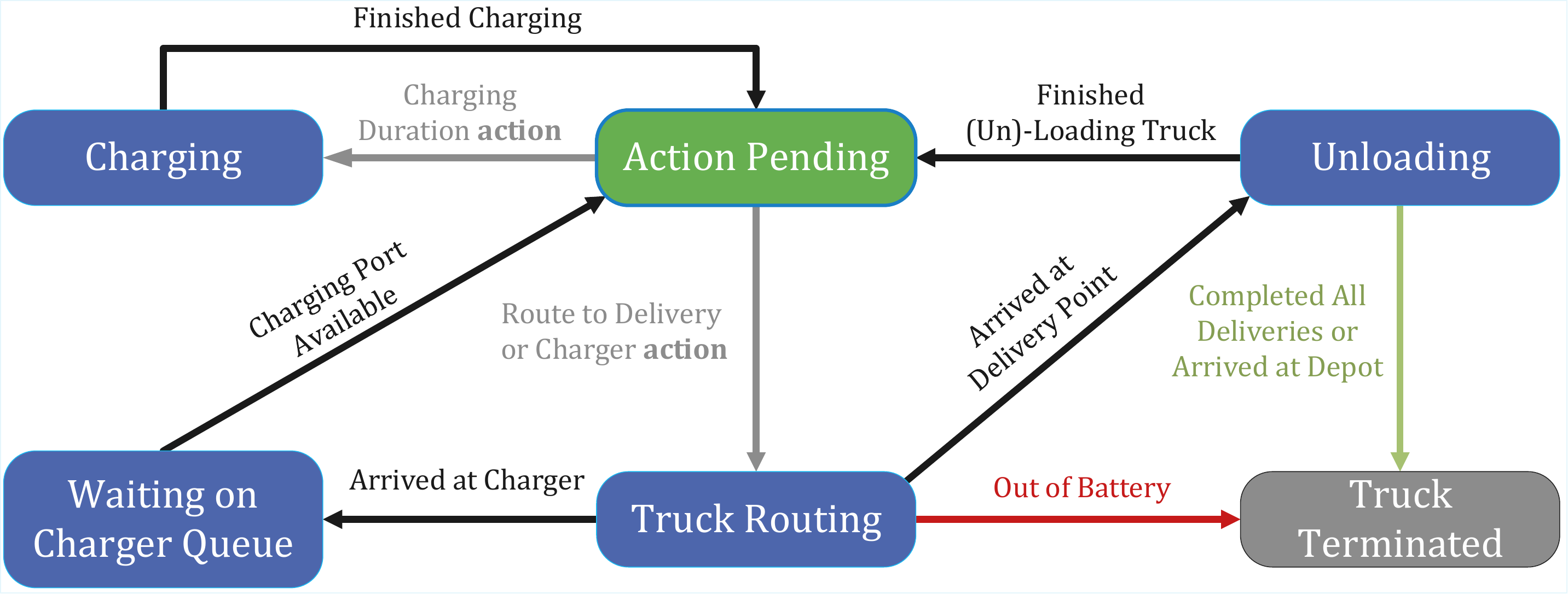}
    \caption{Event-driven truck state machine used by the simulator. Decision steps occur when a truck becomes \texttt{ready}; arrivals, charging, unloading, and FCFS (first-come, first-served) queuing trigger subsequent events.}
    \label{fig:truck_state_machine}
\end{figure}

Each truck transitions among a finite set of operational states that define when it can accept a new decision and when it is constrained by ongoing activities.
In particular, a truck may be in the states: \texttt{Action\_Pending} (available to act), \texttt{Routing} (travel in progress), \texttt{Waiting\_on\_Charger\_Queue} (queued at a station), \texttt{Charging} (occupying a port), \texttt{Unloading} (service in progress), or \texttt{Terminated}, when all deliveries are complete or the truck is out of battery. 
A decision must be made whenever a truck enters the \texttt{Action Pending} state, which occurs after completing a travel segment, finishing unloading at a delivery, or finishing (and vacating) a charging port. 
The resulting event-driven decision structure for each truck is summarized by the state machine in Figure~\ref{fig:truck_state_machine}.

\subsubsection{Charging model}\label{sec:charging_model}
The simulator implements a realistic battery charging model that better reflects DC fast charging than the linear models commonly used in the literature (e.g.,~\cite{BASSO2022102496,lin2022,11016767}).  
During charging, the battery energy increases according to Equation \eqref{eq:con_energy_update_charge}.
In detail, the function $P_c(\cdot)$ returns the net energy added over $[t,t+h]$ and, for DC fast charging, is designed to capture Constant Current--Constant Voltage (CCCV) dynamics through a state of charge (SoC)-dependent tapered power profile. Let $\overline{p}_c$ and $\underline{p}_c$ denote the station-specific peak and minimum charging powers, with SoC defined as $b/\overline{b}$ taking values in $[0,1]$. 
The CCCV charging power is inspired by~\cite{10225616}, and is modeled as a piecewise function:
\begin{equation}
P_c(\text{SoC})=
\begin{cases}
\overline{p}_c \cdot (0.6+3.0 \cdot \text{SoC}), & \text{SoC}\in[0,0.10],\\
\overline{p}_c \cdot\big(0.9+0.25\,(\text{SoC}-0.10)\big), & \text{SoC}\in(0.10,0.50],\\
\overline{p}_c, & \text{SoC}\in(0.50,0.80],\\
\max\!\Big(\underline{p}_{c},\;\overline{p}_c\big(1-0.6\,p^{1.5}\big)\Big), & \text{SoC}\in(0.80,1.0],\;\; \text{where}\,\, p=\frac{\text{SoC}-0.80}{0.20},
\end{cases}
\label{eq:cccv_profile}
\end{equation}
which produces a near-constant-power region at mid SoC and a taper at high SoC. 
In the simulator, \eqref{eq:con_energy_update_charge} is implemented as a discrete integration with a small step $dt$, iteratively updating $b(t)$ and the corresponding SoC-dependent power over $[t,t+h]$. 

\begin{figure}
    \centering
    \includegraphics[width=0.75\linewidth]{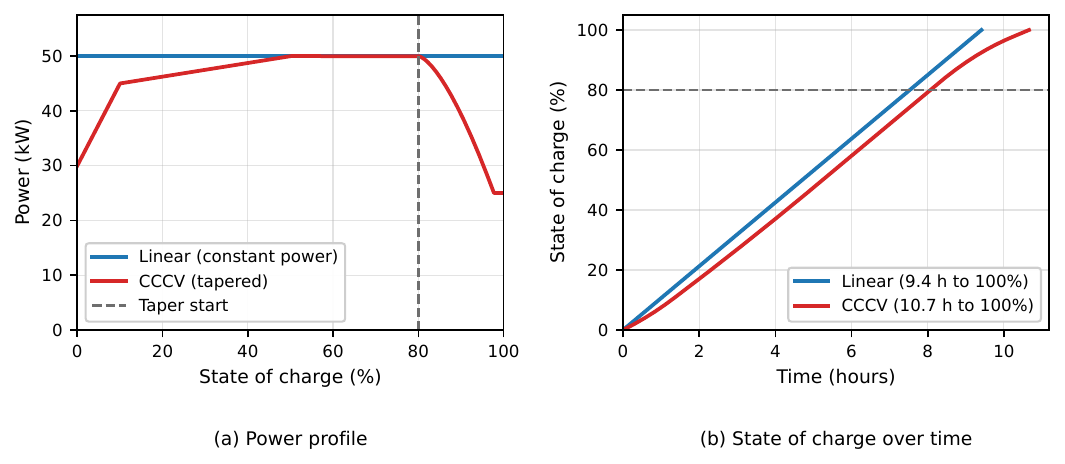}
    \caption{Charging power profile and SoC evolution under constant-power (linear) charging versus tapered CCCV fast charging. Tapering reduces effective power at high SoC and changes optimal charging durations.}
    \label{fig:charging_curve}
\end{figure}

The aforementioned charging model is important for electric truck operations because DC fast charging does not maintain constant power over the full SoC range. Instead, charging power typically tapers as SoC approaches high values, producing diminishing returns in energy gained per unit time. Figure~\ref{fig:charging_curve} visualizes the contrast between a linear charging model and a tapered fast-charging curve, highlighting why decisions about how long to charge cannot be approximated reliably by a linear model at high SoC.


\section{Learning-Based Electric Truck Fleet Routing}

In this section, the eTFRP is formulated as an event-driven semi-Markov Decision Process (SMDP) with a heterogeneous graph state representation and a feasibility-aware, variable-size action representation that mirrors the operational constraints of the truck routing problem. 
Building on this formulation, a graph-based actor-critic architecture and a variable-action GraphPPO training procedure are introduced to enable scalable and generalizable decision-making across instances with varying fleet sizes, delivery sets, and charging infrastructure.

\begin{figure}
    \centering
    \includegraphics[width=0.98\linewidth]{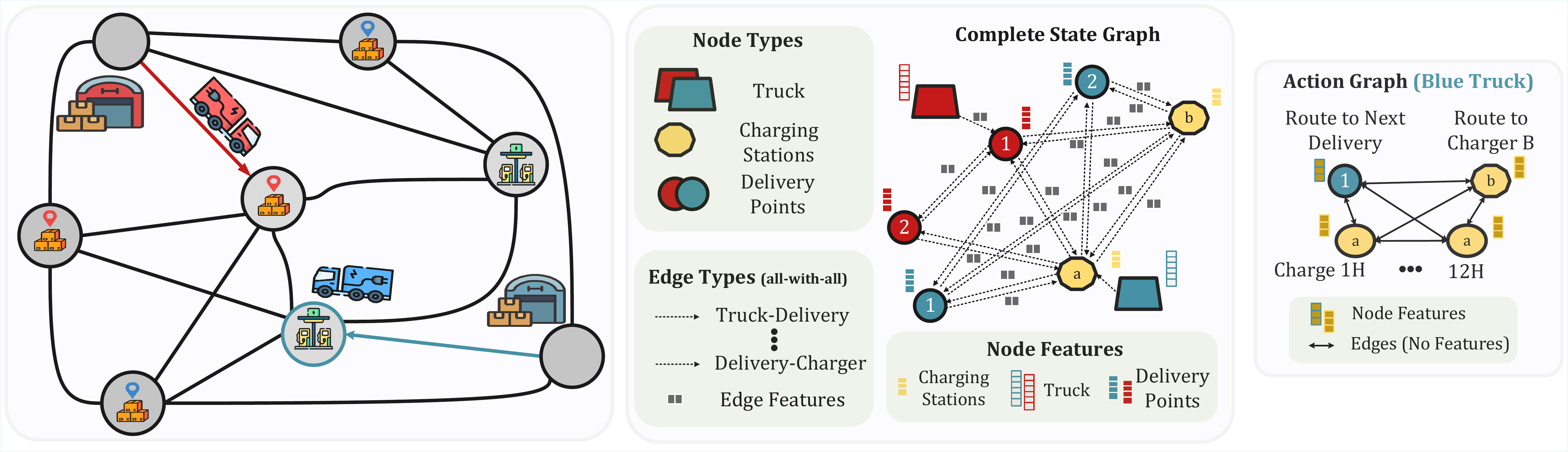}
    \caption{Illustration of the graph construction process at a decision step. The current system state is transformed into a heterogeneous state graph by representing trucks, charging stations, and delivery locations as distinct node types, while feasible routing and charging decisions are encoded in a corresponding action graph.}
    \label{fig:graph_conv}
\end{figure}

\subsection{Graph-Based Semi Markov Decision Process}\label{sec:smdp}
The eTFRP operates in continuous time, and decisions are requested only when a truck enters the \texttt{Action\_Pending} state. 
This event-driven nature of the problem, where each decision step doesn't consist of fixed time intervals, leads to an SMDP~\cite{Elsevier2013_ISBN9780124077959} defined by the tuple
$\big(\mathcal{S},\{\mathcal{A}(s)\}_{s\in\mathcal{S}},P,r,\gamma\big)$,
where $s_t\in\mathcal{S}$ denotes the state observed at the $t$-th decision step, $\mathcal{A}(s_t)$ is the state dependent action space, $P$ is a transition kernel that governs both the next state and the holding time between decisions, $r$ is the reward, and $\gamma\in(0,1)$ is a discount factor. 
The SMDP transition kernel is denoted $P(s_{t+1} \mid s_t, a_t)$ and is determined by the event-driven dynamics introduced in Equations~\eqref{eq:con_battery_bounds}-\eqref{eq:con_ports}. After taking action $a_t$ at state $s_t$, the simulator advances time until the next event (e.g., when a truck is in the ``Action Pending" state) and returns the corresponding next state $s_{t+1}$. 

\subsubsection{The state space}

At each decision step $t$, only a single truck becomes active. If multiple trucks are ready at the exact same time, one is randomly selected; this is rare. 
The state $s_t$ not only contains the state of the active truck $i$, but also contains the full system state needed to evaluate feasibility and to simulate subsequent evolution. The state $s_t\in \mathcal{S}$ is represented as a heterogeneous graph $\mathcal{G}^{\mathcal{S}}_t(\mathcal{V}^{\mathcal{S}}_t,\mathcal{E}^{\mathcal{S}}_t) \subset \mathcal{G}(\mathcal{V},\mathcal{E})$ built on the same entities and symbols used in the problem formulation described in Section~\ref{sec:problem_definition}.
Specifically, as illustrated in Figure~\ref{fig:graph_conv}, a snapshot of the whole eTFRP system state can be converted into the state graph $\mathcal{G}^{\mathcal{S}}_t$ containing three node types comprising the node set $\mathcal{V}^{\mathcal{S}}_t$. Truck nodes are indexed by $i\in\mathcal{I}$, delivery nodes indexed by $d\in\mathcal{D}$, and charging-station nodes indexed by $c\in\mathcal{C}$.

Each truck node has a feature vector $x^{\mathrm{Truck}}_i(t)\in\mathbb{R}$ defined as
\begin{equation}
x^{\mathrm{Truck}}_i(t)\;=\;
\Big[
\frac{b_i(t)}{\overline{b}_i},\;
\text{status}_i(t),\;
\frac{|\mathcal{D}^*_i(t)|}{K_i},\;\
\frac{t_i^{\text{elapsed}}(t)}{T},\;
\frac{t_i^{\text{arrival}}(t)}{T}
\Big]^\top,
\;\;\;\quad \forall i \in \mathcal{I},
\label{eq:truck_features}
\end{equation}
where the first term is the SoC of the truck, the second term denotes the operational status of truck $i$ in the event-driven simulator with a normalized one-hot encoding (e.g., \texttt{ready}, \texttt{routing}, \texttt{charging}, \texttt{waiting\_to\_charge}, \texttt{unloading}).
The set $|\mathcal{D}^*_i(t)| \in \{1,\dots, {K_i}\}$ denotes the number of remaining deliveries for truck $i$, while $K_i$ is the total number of deliveries; hence, the third term is the normalized number of remaining deliveries. The scalar $t_i^{\text{elapsed}}(t)$ denotes the elapsed time since the start of truck $i$'s episode (normalized by the maximum simulation time $T$).
Finally, $t_i^{\text{arrival}}(t)$ shows the estimation of the next ``Action Pending'' event of truck $i$, so it would be zero for the active truck. 

Each pending delivery node for truck $i$ at step $t$ ($d\in\mathcal{D}^*_i(t) \subseteq \mathcal{D}_i$) is associated with feature vector
\begin{equation}
x^{\mathrm{Delivery}}_d(t)\;=\;
\Big[
d_{i,k},\;
|\mathcal{D}^{*}_{i}(t)|,\;
\Big]^\top,
\;\;\;\quad \forall i \in \mathcal{I}, \; \forall d_i \in \mathcal{D}_{i}^*(t).
\label{eq:delivery_features}
\end{equation}
Note that already completed deliveries no longer need to be included in the state space as nodes, as they no longer affect future decisions.

Finally, each charging-station node $c\in\mathcal{C}$ has a feature vector defined as:
\begin{equation}
x^{\mathrm{Charger}}_c(t)\;=\;
\Big[
c,
\overline{p}_c,\;
\eta_c,\;
\kappa_c,\;
\frac{|\mathcal{M}_c(t)|}{\kappa_c},\;
q_c(t)
\Big]^\top,
\forall c \in \mathcal{C}
\label{eq:charger_features}
\end{equation}
where $\overline{p}_c$ is the station peak power, $\eta_c$ the charging efficiency, $\kappa_c$ the number of charging ports, $\mathcal{M}_c(t)\subseteq\mathcal{I}$ the set of trucks occupying ports at time $t$ (as in~\eqref{eq:con_ports}), and $q_c(t)$ denotes the charger queue length at station $c$ at time $t$. Charging prices are assumed to be homogeneous across the traffic network; therefore, they are excluded from the charger state vector.

The edges $\mathcal{E}_t$ of the graph are designed to reflect the transportation costs between all available nodes, as shown in the state graph of Figure~\ref{fig:graph_conv}.
Edges are constructed for each pair of nodes $v,u\in\mathcal{C}\cup\mathcal{D}^*(t)\cup\mathcal{I}$, each with features:
\begin{equation}
x^{\mathrm{Edge}}_{uv}(t)\;=\;\big[\tau_{uv},\; e_{uv}\big]^\top.
\label{eq:edge_features}
\end{equation}
Since $\mathcal{G}$ is not assumed symmetric, these edge features generally satisfy $x^{\mathrm{edge}}_{uv}(t)\neq x^{\mathrm{edge}}_{vu}(t)$ when reversing direction, capturing direction-dependent travel and energy costs.

\subsubsection{The feasible action space}\label{sec:feasible_actions}

The action space in eTFRP is state-dependent because at each decision step only a subset of routing and charging decisions is feasible given the active truck's location, its available battery energy, the set of remaining deliveries, and the instantaneous availability of charging ports. 

To reduce the number of available action combinations, actions that navigate trucks to chargers are limited using a detour-based screening heuristic. 
Let $\bar{d}_{i}(t)$ denote a reference delivery used to measure detours, defined as $\bar{d}_{i}(t)=d_{i}^{k_{i}(t)+1}$ in sequential mode and as the closest remaining delivery in flexible mode.
For each charging station $c\in\mathcal{C}$, the incremental detour (in nominal travel time) of visiting $c$ before continuing toward $\bar{d}_{i_t}(t)$ is computed as
\begin{equation}
\Delta \tau _{i,c} \;=\;\tau_{n_{i}(t),c}+\tau_{c,\bar{d}_{i}(t)}-\tau_{n_{i}(t),\bar{d}_{i}(t)}.
\label{eq:charger_detour}
\end{equation}
Only the $k^{\mathrm{chg}}$ chargers with the smallest detour durations are retained as candidates.
The resulting set of candidate navigation destinations is then
$\hat{\mathcal{V}}_{t}:\{\mathcal{C}^{K_{\mathrm{chg}}}_{i}(t)\ \cup\ \mathcal{D}^{*}_{i}(t)$\}.
This screening step focuses decisions on chargers that are nearby in terms of incremental route deviation, which is particularly relevant in freight operations, where large detours are unlikely to be attractive unless necessitated by feasibility constraints.
Furthermore, a navigation decision to $v\in\hat{\mathcal{V}}_{t}$ is feasible only if it satisfies the conservative energy condition introduced in~\eqref{eq:con_energy_headroom}. This yields the feasible navigation action set
\begin{equation}
\mathcal{A}^{\mathrm{nav}}(s_t)\;=\;\Big\{\;\text{navigate}(v)\;:\; v\in \hat{\mathcal{V}}_{t},\ \alpha\,e_{n_{i}(t),v}<b_{i}(t)\;\Big\},
\label{eq:feasible_nav}
\end{equation}
which filters candidate destinations by requiring sufficient battery energy under the safety multiplier $\alpha\ge 1$.

Charging decisions are only meaningful when the active truck is at a charging station and can initiate charging.
$\mathcal{M}_c(t)\subseteq\mathcal{I}$ is the set of trucks occupying ports at station $c$ at time $t$, and recall that at most $\kappa_c$ trucks can charge simultaneously.
When $n_i(t)=c\in\mathcal{C}$ and a port is available, the truck may select a charging duration $h$ from a discrete set $\mathcal{H}$, which is used to model operationally realistic charging decisions at coarse time intervals. In this work, $\mathcal{H}=\{1,2,\ldots,12\}$ (hours), since a truck should not require more than 12 hours to fully charge, considering fast chargers. The feasible charging actions are therefore
\begin{equation}
\mathcal{A}^{\mathrm{chg}}(s_t)\;=\;\Big\{\;\text{charge}(c,h)\;:\; c=n_{i}(t)\in\mathcal{C},\ |\mathcal{M}_c(t)|<\kappa_c,\ h\in\mathcal{H}\;\Big\}.
\label{eq:feasible_charge}
\end{equation}
The overall feasible action set at step $t$ is given by
$\mathcal{A}(s_t)\;=\;\mathcal{A}^{\mathrm{nav}}(s_t)\ \cup\ \mathcal{A}^{\mathrm{chg}}(s_t)$,
and can equivalently be represented through a binary feasibility mask $m(s_t)$.

The cardinality of $\mathcal{A}(s_t)$ varies across decision steps because both $\mathcal{D}^{*}_{i}(t)$ and the feasibility condition in~\eqref{eq:feasible_nav} depend on the active truck's state, and because charging actions are available only when $n_i(t)\in\mathcal{C}$ and station capacity remains.
For algorithms that require a fixed action dimension (i.e., most RL implementations), a natural upper bound is obtained by enumerating all destination-based navigation actions and all charging-duration actions, yielding a fixed action set size of
$ \lceil\mathcal{A}\rceil \;=\;|\mathcal{C}| \;+\; |\mathcal{D}_{i}(t)| \;+\; |\mathcal{H}|$,
with infeasible actions suppressed at runtime via the binary mask $m(s_t)$.

As shown in Figure~\ref{fig:graph_conv}, at each step $t$, the action set $\mathcal{A}(s_t)$ is also encoded as an action graph $\mathcal{G}^{\mathcal{A}}_t=(\mathcal{V}^{\mathcal{A}}_t,\mathcal{E}^{\mathcal{A}}_t)$,
where $\mathcal{V}^{\mathcal{A}}_t$ contains one node per feasible action. The graph is fully connected with no edge features since the edges do not have a physical meaning, unlike the state graph, where edges represent physical trip distances. 
Each action node $a$ has a three-dimensional feature vector
\begin{equation}
x^{\text{Action}}(t)=\big[\ \mathrm{type}(a),\ \widehat{b}(a),\ \widehat{t}(a)\ \big]^\top,
\label{eq:action_node_features}
\end{equation}
where $\mathrm{type}(a)$ identifies the action category (navigate-to-delivery, navigate-to-charger, or charge),
$\widehat{b}(a)$ is the estimated battery energy after executing $a$ using the nominal updates in \eqref{eq:con_energy_update_travel}--\eqref{eq:con_energy_update_charge}, and
$\widehat{t}(a)$ is the estimated completion time based on the nominal duration of the routing or charging action. 
A graph-based representation of the state and the feasible action set enables GNN policies to exploit network structure while restricting decisions to feasible actions, thereby improving sample efficiency and generalization across varying fleet and infrastructure sizes.

\subsubsection{Reward function}\label{sec:reward}
A time-based reward is adopted to align the learning with the eTFRP objective function in~\eqref{eq:obj_time}, while providing additional shaping signals that encourage task completion and discourage unsafe behavior.
The reward is defined as:
\begin{equation}
r(s_t,a_t,s_{t+1})
=
-\lambda_1\,\Delta t_i
+\lambda_2\,\{\text{delivery completed at }t\}
+\;\lambda_3\,\{\text{failure at }t\},
\label{eq:reward}
\end{equation}
where $\Delta t_i$ is the time passed since the last ``action pending'' status of truck $i$, including stochastic edge traversing, unloading, charging, and waiting at a charger queue times.
This choice ensures that the reward directly reflects the operational time consumed by routing, service, and charging decisions under uncertainty. 
Furthermore, $\lambda_1>0$ and $\lambda_2>0$ are scalar coefficients, and $\lambda_3<0$ is a terminal penalty.
The delivery completion term $\lambda_2$ provides a sparse positive signal whenever a truck successfully completes a delivery, improving exploration and encouraging policies that complete assigned tasks efficiently.
The failure term $\lambda_3$ is applied when the battery is depleted during a routing segment or when the truck becomes stranded without a feasible continuation. The terminal penalty is set to a sufficiently large magnitude to discourage risky decisions that violate energy feasibility, complementing the conservative feasibility logic used in the action set construction.

\subsection{Graph Actor-Critic Network Architecture}\label{sec:gnn_arch}

\begin{figure}
    \centering
    \includegraphics[width=0.8\linewidth]{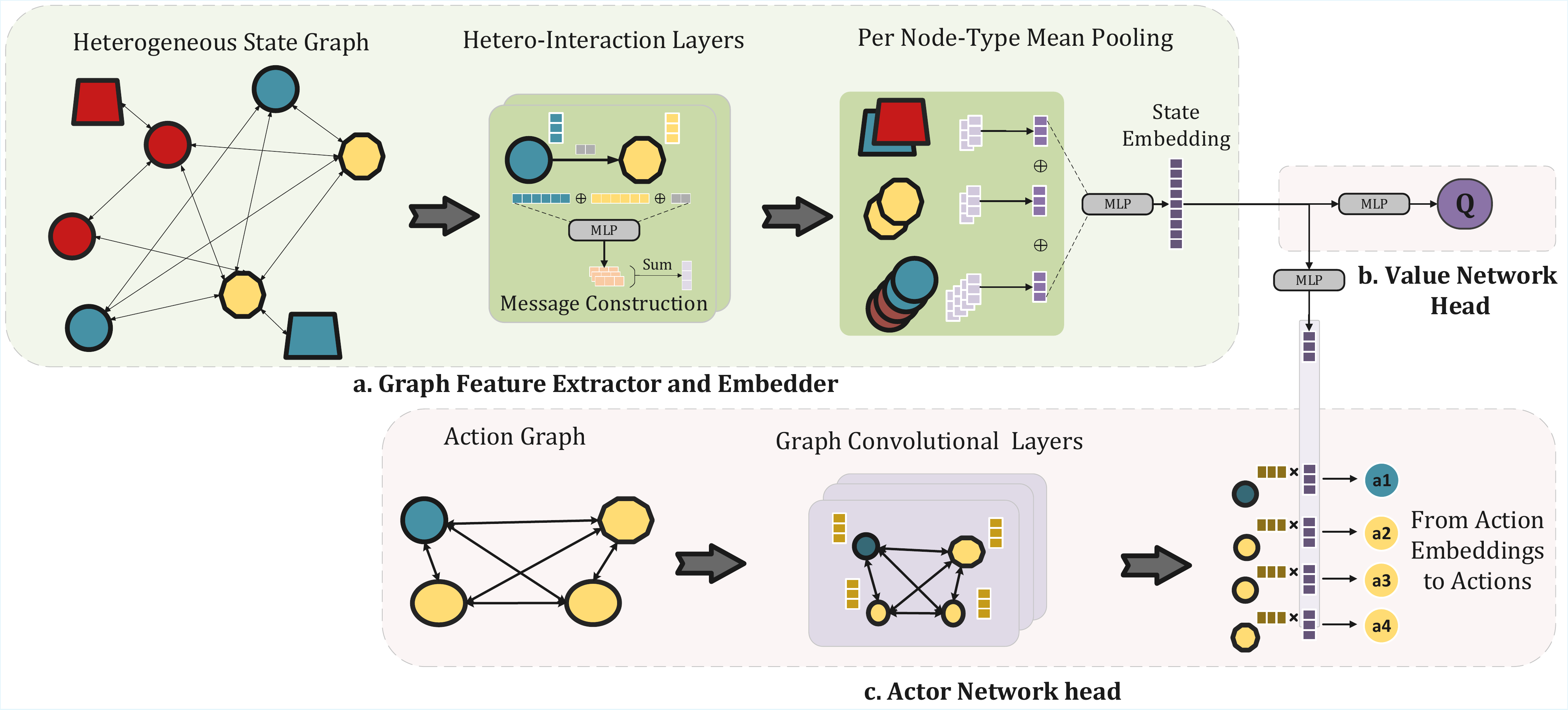}
    \caption{Proposed graph-based actor--critic architecture. In (a), the heterogeneous state graph is encoded through graph feature extraction, hetero-interaction layers~\cite{arowolo2025generalizationgraphneuralnetworks}, and per-node-type mean pooling to obtain a fixed-size global state embedding, invariant to the number of nodes in the state. In (b), the critic head maps this state embedding to a state-value estimate. In (c), the actor head processes the feasible-action graph through graph convolutional layers and combines the resulting action embeddings with the state embedding to produce a policy over routing and charging actions.}
    \label{fig:graph}
\end{figure}

Representing both the system state and the feasible action set as graphs enables an actor--critic architecture that exploits the relational structure of the transportation network and the shared charging infrastructure, while remaining invariant to the number of trucks, deliveries, and charging stations.
The proposed actor and critic networks, visualized in Figure~\ref{fig:graph}, consist of a shared heterogeneous graph encoder that maps the state graph $\mathcal{G}^{\mathcal{S}}_t$ to latent representations, followed by (i) an actor head that scores the nodes of the action graph $\mathcal{G}_t^{\mathcal{A}}$ and creates a categorical policy over $\mathcal{A}(s_t)$, and (ii) a critic head that estimates the state value from a state embedding representation of $\mathcal{G}^{\mathcal{S}}_t$.

\subsubsection{State Graph Encoder}

Initially, the heterogeneous state graph $\mathcal{G}^{\mathcal{S}}_t$ is passed through node specific encoders $f_j(x^j)$ producing node embeddings $\{h_j^{(l)}\}$ for each layer $l$ where $j\in\{\mathrm{Truck},\mathrm{Delivery},\mathrm{Charger}\}$ denotes the node type (as defined in Section~\ref{sec:smdp}).
Each node encoder starts with a single-layer Multilayer Perceptron (MLP), defined as $h_j^{(l)}(x)=W^{(l)}_{j}x+b^{(l)}_{j}$ with the trainable matrix $W^{(l)}\in\mathbf{R}^{F_{l-1} \times F_{l}}$ and bias vector $b^{(l)}\in\mathbf{R}^{F_{l}}$, where $F_{l}$ is the number of hidden nodes.
The node embeddings are then refined through $L$ graph \emph{Hetero-Interaction} layers~\cite{arowolo2025generalizationgraphneuralnetworks}, which perform edge-relation-specific message passing across the directed edges of $\mathcal{G}^\mathcal{S}_t$. 
The set of edges $\mathcal{E}^\mathcal{S}_t$ includes different edge types, denoted by $\mathcal{R}$, (e.g., truck$\rightarrow$delivery, truck$\rightarrow$charger, delivery$\rightarrow$truck, charger$\rightarrow$truck, etc.) where $\mathcal{N}_r(v)$ denotes the neighbors of node $v$ under relation $r\in\mathcal{R}$. 
At layer $l\in\{0,\dots,L-1\}$, the messages of the graph convolution operation~\cite{kipf} are computed with a relation-specific function that conditions on both the sender $u$ and receiver $v$ node embeddings, and the edge feature vector $x^\text{edge}_{uv}$. 
A message of relation $r$ is then defined as:
\begin{equation}
\mu^{(l)}_{uv,r} \;=\; W^{(l)}_{r} \cdot (h_v^{(l-1)} \; \oplus \; h_v^{(l-1)} \; \oplus\; x^\text{edge}_{uv}) +b^{(l)}_{r}.
\label{eq:message}
\end{equation}
The node embedding is then updated by averaging over the relation-specific aggregated messages and combining them with a self-term,
\begin{equation}
h_v^{(l+1)}
\;=\;
\sigma\!\left(
W^{(l)}_{0}\,h_v^{(l)}
\;+\;
\sum_{r\in\mathcal{R}} 
\Big(W^{(l)}_{r} \;
\frac{1}{|\mathcal{N}_r(v)|} \; \sum_{u\in\mathcal{N}_r(v)} \mu^{(l)}_{uv,r}
\Big)
\right).
\label{eq:update}
\end{equation}
where $\sigma(\cdot)$ is a nonlinearity, such as the Rectified Linear Unit (ReLU) function. This construction allows different interaction patterns to be learned for each relation type while preserving permutation invariance within each neighborhood. In particular, directed transportation edges contribute messages that depend on $(\tau_{uv},e_{uv})$, enabling the embeddings to reflect both network topology and traversal costs, whereas interaction edges between trucks and infrastructure nodes propagate fleet-level context such as station congestion and remaining delivery structure.

Finally, as shown in Figure~\ref{fig:graph}.a, the global state embedding $z_t$ is computed by pooling over the node embeddings. In particular, mean pooling is applied per node type to get the intermediate embeddings $z_{j,t}$.
\begin{equation}
z_{j,t}
\;=\;
\frac{1}{\lvert \mathcal{V}^{\mathcal{S}}_{j,t}\rvert}
\sum_{v\in \mathcal{V}_j(\mathcal{G}^{\mathcal{S}}_t)} h_v^{(L)},
\qquad \forall j\in\{\mathrm{Truck},\mathrm{Delivery},\mathrm{Charger}\},
\label{eq:mean_pool_type}
\end{equation}
 Then the intermediate embeddings are concatenated
$z_t \;=\; z_{\mathrm{Truck},t} \;\oplus\; z_{\mathrm{Delivery},t} \;\oplus\; z_{\mathrm{Charger},t}$
to create the final state embedding $z_t$ preserving type information while remaining invariant to the number of nodes in the state graph, such as how many deliveries are left or how many trucks are in the graph.

\subsubsection{Action and Value Function Networks}

As shown in Figure~\ref{fig:graph}.b, the value network is using the state embedding $z_t$, followed by a fully connected layer
$V_{\phi}(s_t)\;=\;\sigma\!\left(W_v \; z_t + b_v\right)$, parameterized by parameters $\phi$.
This value estimate serves as a low-variance baseline for the actor-critic updates, stabilizing learning by reducing the variance of the policy gradient signal.

The actor operates on the feasible-action graph $\mathcal{G}_t^\mathcal{A}({\mathcal{V}^\mathcal{A}},\mathcal{E}_t^{\mathcal{A}})$, which contains one node per feasible action and is fully connected with no edge features.
As shown in Figure~\ref{fig:graph}.c, each action $a\in\mathcal{A}(s_t)$ corresponds to an action node $a\in\mathcal{V}^{\mathcal{A}}$ 
that is refined via $L^{\mathcal{A}}$ layers of graph convolution on the fully-connected feasible-action graph $\mathcal{G}_t^{\mathcal{A}}$. At layer $l\in\{0,\dots,L^{\mathcal{A}}-1\}$, the GCN update for each action node $a$ is
\begin{equation}
h_{a}^{(l+1)}
\;=\;
\sigma\!\left(
W_{\mathrm{a}}^{(l)}
\sum_{v\in\mathcal{N}(a)\cup\{a\}}
\frac{h_{a}^{(l)}}{\sqrt{\tilde{d}_a\,\tilde{d}_v}}
\right),
\label{eq:action_gcn}
\end{equation}
where $\mathcal{N}(a)$ denotes the neighbors of $a$ in $\mathcal{G}_t^{\mathcal{A}}$, $\tilde{d}_a$ and $\tilde{d}_u$ are the degrees of nodes $a$ and $u$ in the graph with self-loops, respectively, and $W_{\mathrm{a}}^{(l)}$ is a learnable weight matrix.
Then, actions are computed by conditioning the action embeddings on the global state embedding $z_t$, in particular by using the dot product $l_t(a) = <h_{a}^{L^{\mathcal{A}}}, z_t>, \; \forall  a\in\mathcal{V}^{\mathcal{A}}$.
The policy is obtained by a softmax over the feasible actions,
\begin{equation}
\pi_{\theta}(a\mid s_t)\;=\;\frac{\exp(l_t(a))}{\sum_{a\in\mathcal{V}^{\mathcal{A}}}\exp(l_t(a))},
\label{eq:actor_policy}
\end{equation}
where $\theta$ is assumed to be all the trainable parameters contributing to the actor network, such as the weights and matrices mentioned before.

Conditioning the actions only on the active-truck embedding $h_{i}^{(L)}$ would have made the estimate sensitive to the local operational context of the currently controlled truck (e.g., battery state, remaining deliveries, and current status), whereas the global state embedding~\cite{orfanoudakis2026gnndt} $z_t$ summarizes system-level interactions and constraints (e.g., shared charger occupancy, congestion, and queueing conditions).
Overall, this actor-critic design uses the state graph encoder to capture system-wide coupling and feasibility structure, while the action graph encoder restricts decision-making to feasible, operationally relevant actions at each event-driven step.

\subsection{Variable Action Handling with GraphPPO}\label{sec:ppo}
Here we introduce GraphPPO (outlined in Algorithm~\ref{alg:vappo}), a PPO~\cite{ppo} variant that uses the graph-based actor--critic architecture of Section~\ref{sec:gnn_arch}.
At each step $t$, the simulator returns the state graph $\mathcal{G}^{\mathcal{S}}_t$ and the feasible action set $\mathcal{A}(s_t)$, from which we build the action graph $\mathcal{G}^{\mathcal{A}}_t$. The actor scores the action nodes to obtain $\pi_\theta(\cdot\mid s_t)$, assigning probability mass only to feasible actions.

The overall objective of RL~\cite{SuttonReinforcementIntroduction} is to find a stochastic policy $\pi_\theta(a\mid s)$ that maximizes the expected discounted return $r$ in the eTFRP, defined as
$J(\theta)\;\triangleq\;\mathbb{E}_{\pi_\theta}\!\left[\sum_{t=0}^{\infty}\gamma^{t}\,r_t\right]$, where $\gamma\in(0,1)$ is the discount factor. 
The critic network is trained using the one-step Bellman target $y_t \;=\; r_t + \gamma\,V_\phi(s_{t+1})$, defined as
\begin{equation}
\mathcal{L}_V(\phi)\;=\;\mathbb{E}_t\!\left[\big(V_\phi(s_t)-y_t\big)^2\right].
\label{eq:bellman_value}
\end{equation}
Using the standard PPO algorithm the advantages are estimated with generalized advantage estimation (GAE),
\begin{equation}
\widehat{A}_t 
\;=\;
\sum_{h=0}^{H-1}(\gamma\lambda)^h\,\big( r_{t+h} + \gamma\,V_\phi(s_{t+h+1}) - V_\phi(s_{t+h}) \big),
\label{eq:gae}
\end{equation}
where $\lambda\in[0,1]$ controls the bias--variance trade-off and $H$ is the rollout horizon.
Let $\theta_{\mathrm{old}}$ denote the parameters used to collect the rollout, and define the importance ratio
\begin{equation}
\rho_t(\theta)\;=\;\frac{\pi_\theta(a_t\mid s_t)}{\pi_{\theta_{\mathrm{old}}}(a_t\mid s_t)}.
\label{eq:ratio}
\end{equation}
GraphPPO then maximizes the clipped surrogate objective
\begin{equation}
\mathcal{L}^{\mathrm{PPO}}(\theta)
=\mathbb{E}_t\Big[
\min\big(\rho_t(\theta)\widehat{A}_t,\ \mathrm{clip}(\rho_t(\theta),1-\epsilon,1+\epsilon)\widehat{A}_t\big)
\Big],
\label{eq:ppo_objective}
\end{equation}
which prevents overly large policy updates by constraining $\rho_t(\theta)$ to a trust region of width $\epsilon$. In practice, we optimize the combined loss consisting of the negative surrogate objective, the value regression loss~\eqref{eq:bellman_value}, and an entropy bonus to encourage exploration.

\begin{algorithm}[t]
\caption{GraphPPO}
\label{alg:vappo}
\begin{algorithmic}[1]
\State \textbf{Input:} discount $\gamma$, clip $\epsilon$, GAE $\lambda$, rollout length $L$, update epochs $E$
\State Initialize actor $\pi_\theta$ and critic $V_\phi$
\For{iteration $=1,2,\dots$}
    \State Set $\theta_{\mathrm{old}}\leftarrow \theta$; clear buffer $\mathcal{B}$
    \For{$t=0,\dots,L-1$}
        \State Observe state $s_t$ and feasible set $\mathcal{A}(s_t)$; build graphs $\mathcal{G}^{\mathcal{S}}_t$ and $\mathcal{G}^{\mathcal{A}}_t$
        \State Compute $\pi_{\theta_{\mathrm{old}}}(\cdot\mid s_t)$; sample $a_t\sim\pi_{\theta_{\mathrm{old}}}(\cdot\mid s_t)$; step env $\rightarrow (r_t,s_{t+1})$
        \State Store $(s_t,a_t,r_t,s_{t+1},\log\pi_{\theta_{\mathrm{old}}}(a_t\mid s_t))$ in $\mathcal{B}$
        \If{episode terminates} \textbf{break} \EndIf
    \EndFor
    \State Compute $\widehat{A}_t$ by GAE~\eqref{eq:gae} and value targets $y_t=r_t+\gamma V_\phi(s_{t+1})$
    \For{epoch $=1,\dots,E$}
        \State Sample minibatches from $\mathcal{B}$
        \State Update actor $\theta$ by maximizing PPO objective~\eqref{eq:ppo_objective} (plus entropy bonus)
        \State Update critic $\phi$ by minimizing $\mathbb{E}\big[(V_\phi(s_t)-y_t)^2\big]$
    \EndFor
\EndFor
\end{algorithmic}
\end{algorithm}


\section{Results}

This section presents the experimental evaluation of the proposed electric truck routing framework. First, the simulation setup is described, including the network configuration, stochastic travel and energy models, charging assumptions, and benchmark settings. The main results are then reported, with emphasis on solution quality, operational feasibility, and computational performance across the tested methods and problem instances.

\subsection{Experimental Setup}

\begin{table}[t]
\centering
\caption{Random variables and experimental settings used in the routing environment.}
\label{tab:random_variables}
\small
\begin{tabular}{p{0.31\linewidth} p{0.61\linewidth}}
\toprule
\textbf{Random variable} & \textbf{Value / configuration} \\
\midrule
Travel time $\tilde{\tau}_{uv}(t)$
& $\tilde{\tau}_{uv}(t)\sim 
\mathcal{N}\!\left(\tau_{uv},\sigma_{uv}^{2}(t)\right)$; 
std.\ factor $0.15$, rush-hour multiplier $2.0$ \\

Energy consumption $\tilde{e}_{uv}$
& 
$\xi_{uv}=\operatorname{clip}\!\left(
1+0.5\!\left(\frac{\tilde{\tau}_{uv}}{\tau_{uv}}-1\right)
+\varepsilon_{uv},\,0.90,\,1.20
\right)$ \\

Charging duration $h$
& $h\in\{1,2,\ldots,12\}$ h \\

Charging model
& CCCV enabled; $p_c=50$ kW, $\eta_c=0.85$, taper from SoC $0.8$ \\

Queueing time
& Endogenous FCFS queue waiting time at charging stations \\

Unloading time
& Fixed at $0.2$ h \\

Initial battery
& Full battery; $b_i=400$ kWh \\
\bottomrule
\end{tabular}
\end{table}

The experimental benchmark is constructed on a directed transportation network derived from real California road-network data~\cite{ziyan2025}. The network is represented through precomputed shortest-path travel-time and energy dictionaries combined with charger metadata, resulting in a graph with 258 candidate delivery nodes, 66{,}049 directed edges, 25 charging stations, and 209 total charging ports.
In every experimental run, only the subset of active and feasible edges is considered.
In the main sequential setting (Table~\ref{tab:random_variables}), all vehicles are modeled as homogeneous electric trucks with a battery capacity of 400~kWh, a base speed of 40~km/h, and a full initial state of charge. At each environment reset, benchmark instances are generated online by assigning each truck a random origin and a feasible ordered delivery sequence satisfying the imposed hop-energy constraints. Across all instance families, the charging infrastructure remains fixed, while experimental difficulty is primarily varied through the number of simultaneously operating trucks.

To reflect operational realism, the benchmark integrates both exogenous and endogenous uncertainty. As shown in Section~\ref{sec:unc}, travel times are modeled as Gaussian perturbations around nominal shortest-path values, with time-dependent variance that increases during rush-hour periods, while realized travel durations are clipped to avoid implausible extremes. Energy consumption is also stochastic and correlated with traffic conditions. In the implemented benchmark, realized energy use is bounded between 0.90 and 1.20 times the nominal value, whereas routing feasibility is evaluated using the conservative upper multiplier of 1.20. Charging congestion is modeled endogenously through finite-capacity charging stations operating under first-come, first-served queues, so waiting times emerge from the realized interaction of fleet trajectories rather than from exogenously specified delays. Charging decisions are discretized into 12 admissible durations ranging from 1 to 12~h, and energy replenishment follows a nonlinear CCCV charging model implemented with a numerical integration step $\Delta t = 0.01$~h. 

\begin{figure}
  \centering
  \includegraphics[width=0.75\linewidth]{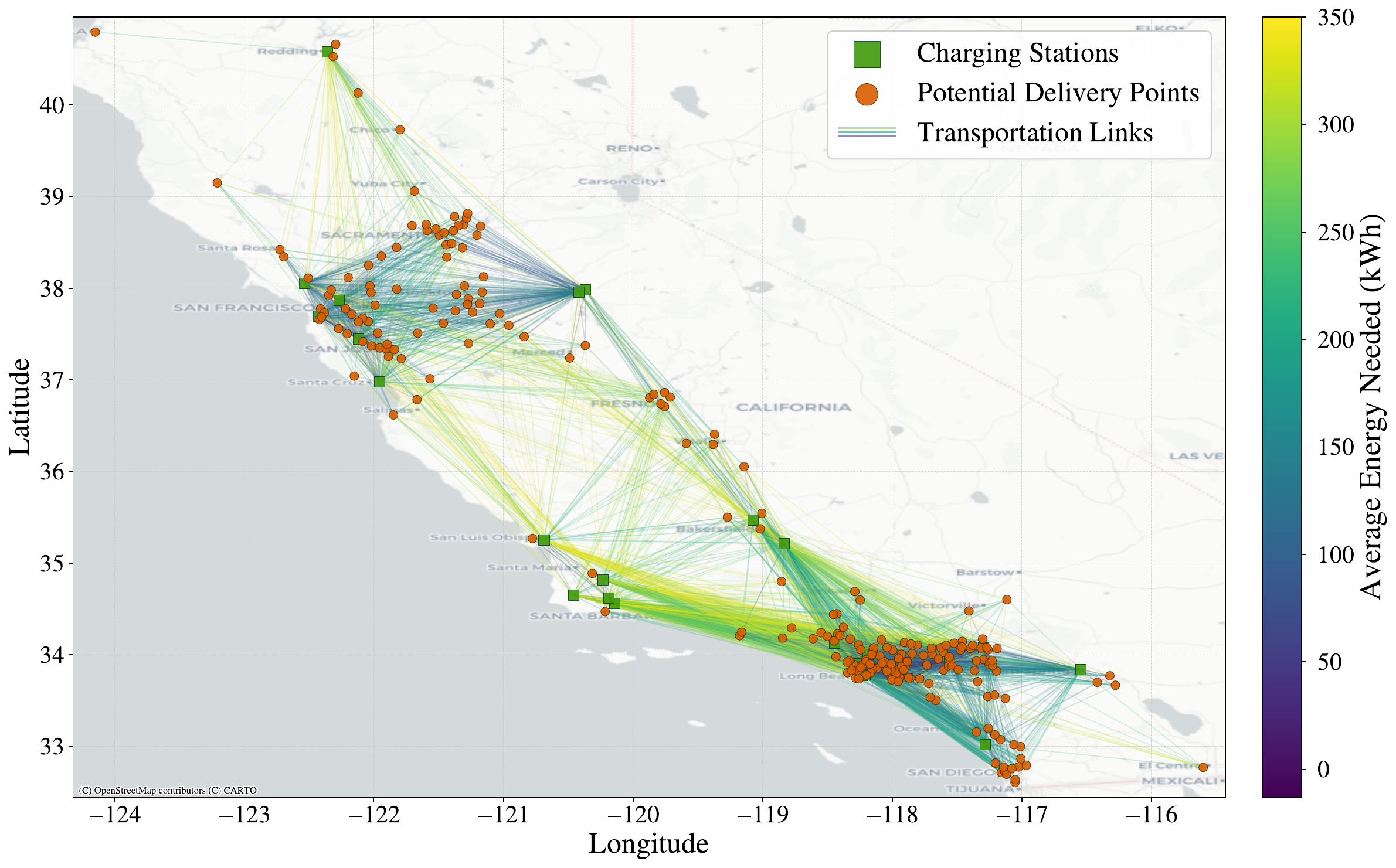}
\caption{Experimental transportation network used in the case study. The figure shows the California transportation graph~\cite{ziyan2025} with sampled delivery locations and charging stations overlaid on the map. Node and edge colors indicate the spatial variation in energy and travel conditions across the network.}
\label{fig:cali}
\end{figure}

All reported results are averages over 200 evaluation scenarios, and training is performed with 5 random seeds to ensure statistical significance. GraphPPO's feature extractor has 3 graph layers, each with 256 nodes, a critic MLP head with 1 layer and 256 nodes, a learning rate of $3\times10^{-4}$, and a discount factor of $\gamma = 0.99$. For both GraphPPO and the PPO-based baselines, the nominal training budget is $10^6$ environment steps with periodic evaluation during training. 
The generic PPO and MaskPPO baselines use the default Stable-Baselines3~\cite{sb3} configuration, with training requiring approximately 5--6 hours, compared with about 10 hours for the proposed GraphPPO. Although training requires several hours offline, the resulting models can generate high-quality solutions in real time at deployment.
Training is done on a system with an A10 GPU and a 16-core Intel CPU, and 24 GB of RAM.

\subsection{Performance Across Truck Fleet Scales}

Five routing algorithms are considered in this comparison. Mathematical optimization is used as a high-quality planning reference, obtained by solving the optimization model conservatively, as shown in Section~\ref{sec:obj_constraints}, under the same environment assumptions, but without fully accounting for the problem's stochastic nature.
The heuristic algorithm is a rule-based baseline designed to construct feasible routing and charging decisions with low computational effort but without learning, generic PPO~\cite{ppo} is the most capable discrete-action RL algorithm and serves as the standard method applied without any explicit structural adaptation to the transportation graph or the feasibility structure of the problem, while MaskPPO~\cite{maskppo} augments PPO with an action-feasibility mask so that invalid routing and charging actions are removed during decision making. GraphPPO further extends this idea through graph-based state and action representations that encode interactions among trucks, deliveries, and charging stations under shared infrastructure constraints.

\begin{figure}
  \centering
  \includegraphics[width=1\linewidth]{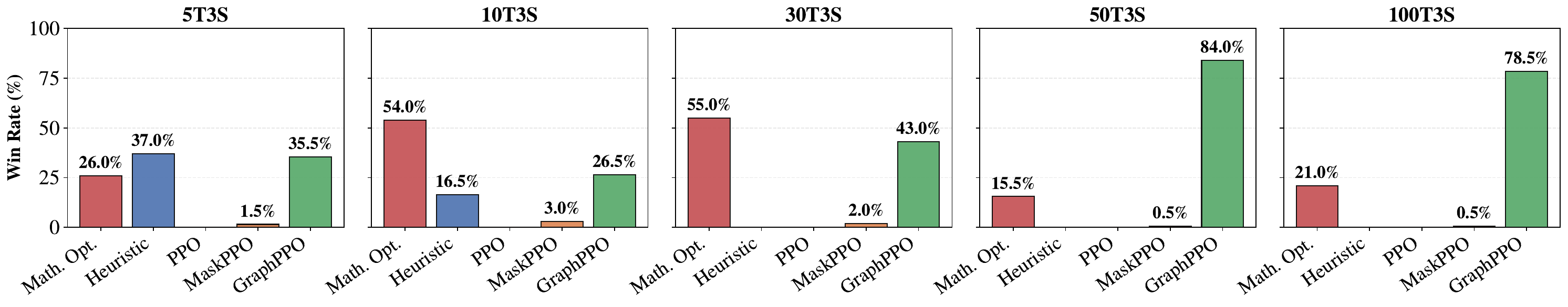}
\caption{Comparison of episode-wise winning frequency across 200 simulation episodes for different eTFRP settings. Each bar shows the percentage of episodes in which a method produced the best-performing solution among the compared methods.}
\label{fig:wr}
\end{figure}

Table~\ref{tab:genresults} summarizes the eTFRP performance across fleet sizes from  1 truck with 3 stops (1T3S) to 100 trucks (100T3S), and Figure~\ref{fig:wr} complements these averages by showing which method achieves the best outcome (referred to as ``win ratio'') on each individual test scenario.
Although each truck completes exactly three delivery stops in this setting, the proposed formulation readily extends to an arbitrary number of stops, as demonstrated in the next section.
Performance is evaluated using normalized reward (relative to the mathematical benchmark) and the scheduling success rate (number of routes completed without any stranded trucks). Therefore, values close to $1.0$ indicate near-benchmark performance, values substantially below $1.0$ indicate a clear performance gap, and values slightly above $1.0$ may still be observed because evaluation is carried out under stochastic simulation over 200 random scenarios, where Math. Opt. provides a conservative solution. Here, all learning algorithms are trained specifically for each problem setting.

A clear separation among methods is reported in Table~\ref{tab:genresults}, with GraphPPO being the only learning-based method for which near-benchmark performance is retained across all settings, since normalized rewards of $0.987$ to $1.005$ are obtained for even the hardest setting with 100 trucks, while the corresponding success rates remain similar to the Math. Opt. solution. In contrast, PPO yields very weak results, with normalized reward ranging from $0.081$ to $0.321$ and success falling to $0\%$ from 5T3S onward. Furthermore, MaskPPO shows the strength of the proposed feasible action mask (shown in Section~\ref{sec:feasible_actions}) with normalized reward from $0.972$ to $0.989$ up to 30T3S, while its robustness is sharply reduced as scale increases, with success declining from $98.0\%$ in 1T3S to $53.0\%$ in 30T3S, $4.5\%$ in 50T3S, and $2.0\%$ in 100T3S. Meanwhile, the heuristic remains competitive only in smaller settings and is reduced to $0\%$ success from 30T3S onward.

\label{sec:general-results}
\begin{table}
\centering
\small
\caption{Performance comparison of optimization, heuristic, and learning-based methods across different eTFRP settings. Each learning-based model was trained specifically for the corresponding eTFRP setting and subsequently evaluated on 200 random scenarios. Entries report normalized reward mean and standard deviation, while the success rate is shown within the parentheses (\%). Bold numbers denote the best normalized reward for each setting.}
\resizebox{1\linewidth}{!}{
\begin{tabular}{@{}lccccc@{}}
\toprule
Setting & Math. Opt. & Heuristic & PPO & MaskPPO & GraphPPO (Ours) \\
\midrule
1T3S & $\mathbf{1.000}$ ($99.5$\%) & $0.806$ $\pm$ $0.448$ ($83.0$\%) & $0.081$ $\pm$ $0.428$ ($15.5$\%) & $0.972$ $\pm$ $0.229$ ($98.0$\%) & $0.987$ $\pm$ $0.136$ ($99.0$\%) \\
5T3S & $\mathbf{1.000}$ ($97.5$\%) & $0.814$ $\pm$ $0.213$ ($42.5$\%) & $0.091$ $\pm$ $0.198$ ($0.0$\%) & $0.980$ $\pm$ $0.085$ ($89.5$\%) & $\mathbf{1.000}$ $\pm$ $0.039$ ($97.0$\%) \\
10T3S & $\mathbf{1.000}$ ($95.5$\%) & $0.803$ $\pm$ $0.150$ ($16.5$\%) & $0.083$ $\pm$ $0.152$ ($0.0$\%) & $0.989$ $\pm$ $0.055$ ($87.5$\%) & $\mathbf{1.000}$ $\pm$ $0.027$ ($95.0$\%) \\
30T3S & $\mathbf{1.000}$ ($88.5$\%) & $0.796$ $\pm$ $0.082$ ($0.0$\%) & $0.125$ $\pm$ $0.086$ ($0.0$\%) & $0.974$ $\pm$ $0.035$ ($53.0$\%) & $0.999$ $\pm$ $0.014$ ($86.5$\%) \\
50T3S & $1.000$ ($77.0$\%) & $0.802$ $\pm$ $0.066$ ($0.0$\%) & $0.305$ $\pm$ $0.090$ ($0.0$\%) & $0.948$ $\pm$ $0.036$ ($4.5$\%) & $\mathbf{1.005}$ $\pm$ $0.013$ ($76.5$\%) \\
100T3S & $1.000$ ($59.5$\%) & $0.800$ $\pm$ $0.048$ ($0.0$\%) & $0.321$ $\pm$ $0.059$ ($0.0$\%) & $0.959$ $\pm$ $0.023$ ($2.0$\%) & $\mathbf{1.005}$ $\pm$ $0.011$ ($60.0$\%) \\
\bottomrule
\end{tabular}
}
\label{tab:genresults}
\end{table}

Additional results are provided in Figure~\ref{fig:wr} through episode-wise win rates over the 200 evaluation scenarios, where in 5T3S the highest share is attained by the heuristic at $37.0\%$, with GraphPPO close behind at $35.5\%$ and Math.\ Opt.\ at $26.0\%$, while in 10T3S and 30T3S the largest win rates are still recorded for Math.\ Opt.\ at $54.0\%$ and $55.0\%$, with GraphPPO remaining second at $26.5\%$ and $43.0\%$, and in the largest settings a clear shift is observed, since GraphPPO attains $84.0\%$ in 50T3S and $78.5\%$ in 100T3S, compared with only $15.5\%$ and $21.0\%$ for Math.\ Opt., from which it can be inferred that near-benchmark average performance is not only preserved by GraphPPO, but superior competitiveness is also achieved in the most congested large-fleet regimes.

\subsection{Generalization from a Single Training Setting}

\begin{figure} 
\centering
     \subfloat[Normalized reward ratio]{
         \centering
         \includegraphics[width=0.4\linewidth]{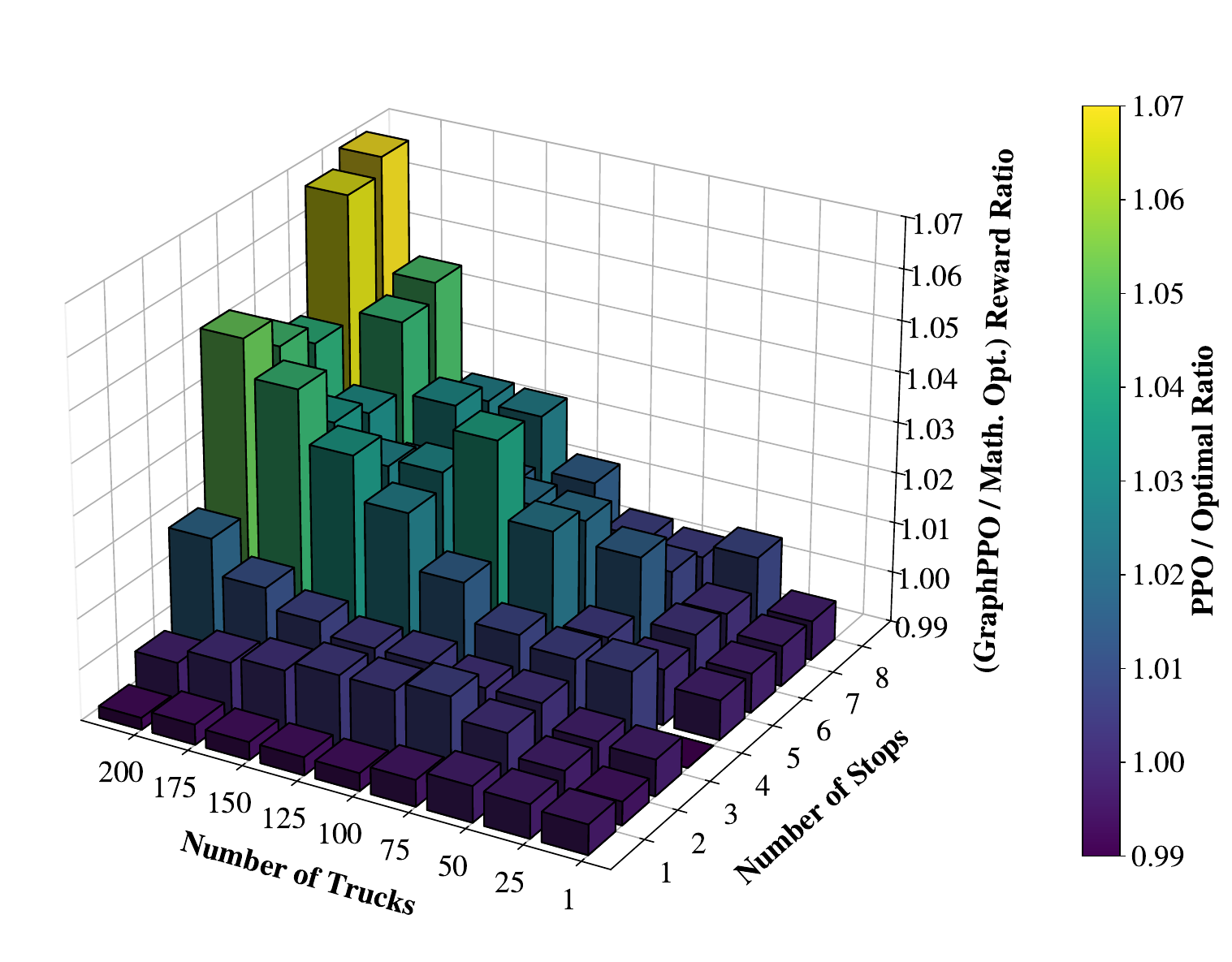}
         \label{fig:gen_a}
         }
        \hspace{10mm}
     \subfloat[Win ratio]{
         \centering
         \includegraphics[width=0.4\linewidth]{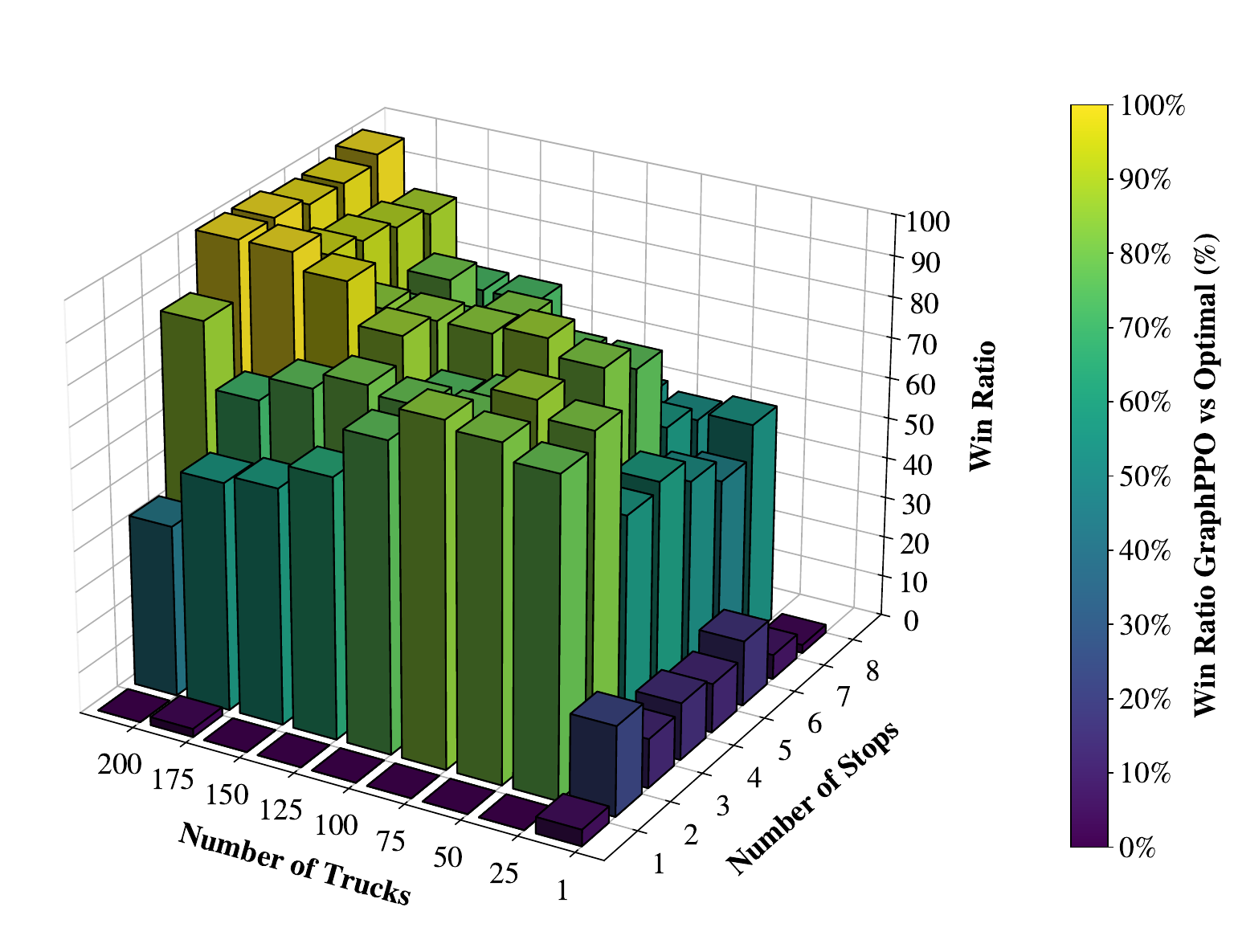}
         \label{fig:gen_b}
         }
    \caption{Generalization performance of a GraphPPO policy trained only on the 100T3S setting and evaluated without retraining across different eTFRP configurations, averaged over 50 random scenarios per case. Panel (a) reports the normalized reward ratio relative to Math.\ Opt., while panel (b) reports the win ratio against Math.\ Opt., defined as the percentage of scenarios in which GraphPPO achieves the better outcome.}
        \label{fig:gen}
\end{figure}

A zero-shot generalization analysis is conducted in Figure~\ref{fig:gen}, where a single GraphPPO policy trained only on the 100T3S setting is applied directly to all other problem configurations without any retraining or fine-tuning. The horizontal axes span the number of trucks and the number of sequential stops per truck, while each reported value is averaged over 50 random scenarios. Two complementary metrics are shown. Panel~(a) reports the normalized reward ratio, defined as the average GraphPPO reward divided by the average reward of Math.\ Opt.\ for the same setting, so values above $1.0$ indicate that the transferred policy outperforms the conservative optimization benchmark on average. Panel~(b) reports the win ratio against Math.\ Opt., i.e., the percentage of scenarios in which GraphPPO achieves a better outcome than the optimization benchmark.

Strong transfer performance is observed over a broad region of the test space. In panel~(a), the normalized reward ratio remains very close to or above parity in almost all settings, typically lying around $1.00$ to $1.04$, while the strongest gains are obtained in larger and more congested instances, where values rise to approximately $1.06$ to $1.07$. The largest advantage is visible in high-truck and mid-to-high stop settings, especially around 150 to 200 trucks and roughly 5 to 7 stops, indicating that the policy learned on 100T3S captures decision patterns that remain effective as congestion and fleet interactions become even stronger.
At the same time, only limited degradation is observed in smaller settings. A likely explanation is that instances with few trucks or stops offer relatively little variability in feasible solutions, making the decision landscape simpler. Consequently, the reward ratio remains close to 1.0, suggesting that the learned policy does not over-specialize to the training configuration.

The win-ratio results in panel~(b) further support this conclusion, although a more heterogeneous pattern is observed because this metric is stricter than average reward comparison. For medium and large fleets with approximately 2 to 6 stops, GraphPPO wins in a large majority of scenarios, with win ratios often between $70\%$ and $100\%$, and with the highest values appearing in the dense multi-truck regimes where charger competition is strongest. By contrast, the smallest instances and the one-stop cases are associated with much lower win ratios, in some cases close to zero, indicating that the optimization benchmark remains more competitive when the problem structure is simple, and congestion at charging stations is limited. Overall, the results indicate that the policy trained on 100T3S generalizes well across both smaller and larger eTFRP settings, with the most substantial advantage retained in the operational regimes most relevant to large-scale fleet coordination.

\subsection{In-Depth Analysis for a Single Setting}

A more detailed comparison is provided through the representative 10T3S setting, which corresponds to 10 trucks and 3 sequential delivery stops per truck, while all reported values in Table~\ref{tab:10t3s_metrics} are averaged over 200 evaluation scenarios and are shown as mean $\pm$ standard deviation.

The 10T3S setting is selected as a representative benchmark because it strikes a balance between problem complexity and computational tractability. With 10 trucks and 3 sequential delivery stops per truck, the setting introduces sufficient combinatorial diversity to meaningfully evaluate the model's decision-making capabilities while remaining interpretable for detailed analysis.
Table~\ref{tab:10t3s_metrics} shows detailed results averaged over 200 evaluation scenarios.
The reported {Reward} corresponds to the episodic return used by the environment, which jointly reflects routing, charging, waiting, unloading, successful delivery, and failure penalties. The {Success Rate} indicates the percentage of scenarios completed feasibly. The {Avg.\ Truck SoC at Finish} measures the average remaining battery level at the end of the episode, while {Total Deliveries} reports how many deliveries are completed across the fleet. The charging-related indicators are separated into {Total Charging Sessions}, which counts the number of charging actions; {Total Charging Time}, which measures the aggregate time spent charging, and {Total Waiting Time}, which captures the time spent in charger queues. In addition, {Total Routing Time} and {Unloading Time} quantify time spent traveling and servicing deliveries, {Total Time} summarizes the overall operational duration, and Exec.\ Time reports the wall-clock solution time required by each method.

\begin{table}
\centering
\small
\caption{Detailed metric performance comparison for the 10T3S setting over 200 evaluation scenarios. Reported values are the mean and standard deviation over the evaluation scenarios.}
\begin{tabular}{@{}lccccc@{}}
\toprule
Metric & Math. Opt. & Heuristic & PPO & MaskPPO & GraphPPO (Ours) \\
\midrule
Reward (-) & $14582$ $\pm$ $583$ & $11718$ $\pm$ $2245$ & $1244$ $\pm$ $2205$ & $14404$ $\pm$ $721$ & $14574$ $\pm$ $568$ \\
Success Rate (\%) & $95.5$ $\pm$ $20.8$ & $16.5$ $\pm$ $37.2$ & $0.0$ $\pm$ $0.0$ & $87.5$ $\pm$ $33.2$ & $95.0$ $\pm$ $21.8$ \\
Avg. Truck SoC at Finish (\%) & $36.0$ $\pm$ $3.4$ & $14.6$ $\pm$ $6.4$ & $27.0$ $\pm$ $6.4$ & $50.1$ $\pm$ $7.1$ & $14.2$ $\pm$ $2.4$ \\
Total Deliveries (-) & $29.9$ $\pm$ $0.7$ & $27.4$ $\pm$ $2.2$ & $19.4$ $\pm$ $2.3$ & $29.7$ $\pm$ $0.9$ & $29.9$ $\pm$ $0.6$ \\
Total Charging Sessions (-) & $13.5$ $\pm$ $1.8$ & $9.2$ $\pm$ $2.0$ & $0.0$ $\pm$ $0.0$ & $11.4$ $\pm$ $5.3$ & $13.5$ $\pm$ $2.6$ \\
\midrule
Total Charging Time (H) & $63.1$ $\pm$ $10.0$ & $28.4$ $\pm$ $7.7$ & $0.0$ $\pm$ $0.0$ & $87.8$ $\pm$ $19.0$ & $48.6$ $\pm$ $12.3$ \\
Total Waiting Time (H) & $4.3$ $\pm$ $8.2$ & $0.7$ $\pm$ $2.0$ & $0.0$ $\pm$ $0.0$ & $5.4$ $\pm$ $9.0$ & $2.0$ $\pm$ $3.7$ \\
Total Routing Time (H) & $224.3$ $\pm$ $16.6$ & $207.8$ $\pm$ $20.3$ & $132.5$ $\pm$ $12.7$ & $246.7$ $\pm$ $29.6$ & $249.6$ $\pm$ $23.2$ \\
Unloading Time (H)& $4.0$ $\pm$ $0.1$ & $3.8$ $\pm$ $0.2$ & $3.5$ $\pm$ $0.3$ & $4.0$ $\pm$ $0.1$ & $4.0$ $\pm$ $0.1$ \\
Total Time (H) & $295.8$ $\pm$ $27.3$ & $240.7$ $\pm$ $26.9$ & $136.0$ $\pm$ $12.9$ & $343.9$ $\pm$ $48.0$ & $304.1$ $\pm$ $35.0$ \\
\midrule
Exec. Time (s) & $1.9$ $\pm$ $0.7$ & $0.1$ $\pm$ $0.1$ & $0.2$ $\pm$ $0.2$ & $1.0$ $\pm$ $0.7$ & $1.7$ $\pm$ $0.3$ \\
\bottomrule
\end{tabular}
\label{tab:10t3s_metrics}
\end{table}

For this case, near-equivalence between GraphPPO and the mathematical benchmark is observed in the two most important outcome metrics. A reward of $14574$ is obtained by GraphPPO, compared with $14582$ for Math.\ Opt., while the success rate remains almost identical at $95.0$ versus $95.5$. A comparable number of deliveries is also completed, with both methods averaging $29.9$ deliveries, and an identical average number of charging sessions is recorded at $13.5$. These results indicate that the proposed policy reproduces the overall quality and feasibility of the optimization benchmark under stochastic execution. By contrast, substantial degradation is observed for the heuristic and PPO baselines. The heuristic reaches a reward of $11718$ with only $16.5\%$ success, while generic PPO collapses completely with $0.0\%$ success and only $19.4$ completed deliveries. MaskPPO remains much stronger than PPO, with a reward of $14404$ and success of $87.5\%$, but still remains below both Math.\ Opt.\ and GraphPPO in reliability.

The operational metrics further clarify the behavioral differences among methods. GraphPPO ends with a lower residual SoC than Math. Opt.\ at $14.2\%$ versus $36.0\%$, which suggests that less conservative battery utilization is being adopted while feasibility is still preserved. This behavior is also reflected in the charging profile, since GraphPPO spends only $48.6$ hours in charging compared with $63.1$ for Math. Opt., and queueing time is reduced from $4.3$ hours to $2.0$ h. At the same time, a longer routing time is observed for GraphPPO at $249.6$ hours versus $224.3$, so the final total time is slightly higher at $304.1$ hours compared with $295.8$. A different pattern is observed for MaskPPO, where a very high terminal SoC of $50.1\%$ is retained together with $87.8$ hours of charging time and $343.9$ hours of total time, indicating an overly conservative charging strategy. Finally, low online solution times are maintained by all methods, with $1.7$ seconds for GraphPPO and $1.9$ seconds for Math.\ Opt. (since there are only 10 trucks), while the heuristic and PPO-based baselines remain below one second, so near-real-time deployment remains feasible for the proposed policy in this setting.

\subsection{From eTFRP to Single-Truck eVRP}

In the previous subsections, performance was assessed for the complex stochastic eTFRP across multiple fleet scales, with routing quality coupled with shared charging infrastructure, queueing effects, and interactions among concurrently operating trucks. To further illustrate the adaptability of the proposed framework and its ability to address a wider range of routing problems, the focus now shifts to the flexible-delivery-order, single-truck, stochastic eVRP setting. In this setting, fleet-level interactions are eliminated, leading to a fundamentally different decision structure.
In the 1T20S case, a single truck must serve 20 delivery locations, but the delivery order is not fixed in advance, so the sequence of customer visits must also be determined as part of the decision process in order to minimize total trip time. As a result, the problem is no longer dominated by charger competition, but instead by the joint difficulty of route construction, battery-feasible navigation, and charging planning over a longer delivery horizon. Since only one truck is considered, waiting time is zero for all methods. The heuristic baseline used in this setting is different from before and is inspired by the TSP. In detail, heuristic baseline is a nearest-neighbor route-construction method followed by local improvement steps, so a feasible tour is first built greedily by repeatedly visiting the closest unserved customer and is then refined through edge swaps that reduce total travel cost~\cite{5ade80e3-7757-300c-a7d3-653ed50f3347}.

\begin{table}
\centering
\small
\caption{Detailed performance comparison for the 1T20S single-truck eVRP setting over 100 feasible evaluation scenarios. Reported values are the mean $\pm$ standard deviation and include operational metrics.}
\begin{tabular}{@{}lccccc@{}}
\toprule
Metric & Math. Opt. & Heuristic & PPO & MaskPPO & GraphPPO (Ours) \\
\midrule
Reward (-) & $10473$ $\pm$ $13$ & $9553$ $\pm$ $1379$ & $9089$ $\pm$ $1811$ & $10401$ $\pm$ $267$ & $10443$ $\pm$ $65$ \\
Success Rate (\%) & $100.0$ $\pm$ $0.0$ & $48.9$ $\pm$ $50.3$ & $45.6$ $\pm$ $50.1$ & $96.7$ $\pm$ $18.1$ & $98.9$ $\pm$ $10.5$ \\
Avg. Truck SoC at Finish (\%) & $28.9$ $\pm$ $6.2$ & $9.3$ $\pm$ $6.4$ & $11.9$ $\pm$ $10.8$ & $55.6$ $\pm$ $25.0$ & $60.7$ $\pm$ $19.1$ \\
Total Deliveries (-) & $20.0$ $\pm$ $0.0$ & $19.1$ $\pm$ $2.2$ & $18.6$ $\pm$ $2.7$ & $20.0$ $\pm$ $0.2$ & $20.0$ $\pm$ $0.1$ \\
Total Charging Sessions (-) & $0.9$ $\pm$ $0.8$ & $10.8$ $\pm$ $16.7$ & $0.0$ $\pm$ $0.0$ & $1.3$ $\pm$ $0.9$ & $2.2$ $\pm$ $6.8$ \\
\midrule
Total Charging Time (H) & $3.1$ $\pm$ $4.8$ & $11.6$ $\pm$ $17.8$ & $0.0$ $\pm$ $0.0$ & $11.9$ $\pm$ $6.7$ & $13.2$ $\pm$ $7.1$ \\
Total Routing Time (H) & $21.7$ $\pm$ $7.8$ & $45.1$ $\pm$ $36.2$ & $20.1$ $\pm$ $2.6$ & $34.8$ $\pm$ $11.0$ & $36.0$ $\pm$ $12.2$ \\
Unloading Time (H) & $2.0$ $\pm$ $0.0$ & $2.0$ $\pm$ $0.2$ & $1.9$ $\pm$ $0.3$ & $2.0$ $\pm$ $0.0$ & $2.0$ $\pm$ $0.0$ \\
Total Time (H) & $26.8$ $\pm$ $12.5$ & $58.6$ $\pm$ $53.9$ & $22.0$ $\pm$ $2.8$ & $48.7$ $\pm$ $17.3$ & $51.2$ $\pm$ $19.1$ \\
\midrule
Exec. Time (s) & $233.1$ $\pm$ $106.4$ & $0.2$ $\pm$ $0.1$ & $0.2$ $\pm$ $0.2$ & $0.2$ $\pm$ $0.2$ & $0.8$ $\pm$ $0.3$ \\
\bottomrule
\end{tabular}
\label{tab:1t20s_metrics}
\end{table}

As observed in Table~\ref{tab:1t20s_metrics}, near-optimal performance is again obtained by GraphPPO, although the relative differences among the strongest methods become smaller in this setting because the optimization benchmark is no longer affected by resource competition. A mean reward of $10443$ is obtained by GraphPPO, compared with $10473$ for Math.\ Opt., while the success rate remains at $98.9\%$ versus $100.0\%$ for the optimization benchmark. Full-service completion is also essentially preserved, with $20.0$ deliveries completed by GraphPPO, compared with $20.0$ by Math.\ Opt. A similarly strong pattern is observed for MaskPPO, which achieves a reward of $10401$ and a success rate of $96.7\%$. In contrast, much weaker results are obtained with the heuristic and generic PPO baselines, whose success rates are only $48.9\%$ and $45.6\%$, respectively, and their rewards are $9553$ and $9089$. These results indicate that, even in the absence of charger congestion, explicit feasibility-aware learning remains important, whereas generic RL without structural adaptation remains unreliable.

The operational metrics show that the remaining gap between GraphPPO and Math.\ Opt.\ is not driven by missed deliveries, but mainly by differences in charging and routing behavior. Math.\ Opt.\ completes the route with only $0.9$ charging sessions on average and $3.1$ hours of charging time, together with the shortest total time of $26.8$ hours. By comparison, GraphPPO achieves nearly the same reward and delivery completion level, but does so with a more conservative energy strategy, finishing with an average SoC of $60.7\%$, using $2.2$ charging sessions, and spending $13.2$ hours charging, for a total time of $51.2$ hours. A similar conservative pattern is observed for MaskPPO, which finishes with $55.6\%$ SoC and a total time of $48.7$ hours. The heuristic, by contrast, is associated with inefficient charging and routing decisions, with $10.8$ charging sessions, $11.6$ hours of charging time, and $58.6$ hours of total time, while PPO fails even more severely, with no charging actions, only $18.6$ deliveries completed, and early infeasible termination. The route visualization supports this interpretation, since the optimization- and feasibility-aware learning methods are associated with coherent long-horizon routing patterns, whereas the weaker baselines are associated with inefficient detours or premature failure. A substantial computational advantage is nevertheless retained by GraphPPO, since Math.\ Opt.\ requires $233.1$ seconds per instance, whereas GraphPPO requires only $0.8$ seconds, which shows that near-optimal eVRP performance can be obtained with much lower online computation.



        
        
        


\section{Conclusion}

This paper introduced a truck-oriented, learning-compatible framework for the eTFRP and eVRP under charging constraints, nonlinear charging behavior, and operational uncertainty. The problem was formulated as an event-driven semi-Markov decision process with shared charging resources, stochastic travel and energy consumption, and feasibility-aware decision-making, and solved using the proposed GraphPPO method. Across the main multi-truck benchmark, GraphPPO remained consistently close to the mathematical programming baseline, achieving normalized rewards of 0.987--1.005 from 1 to 100 truck instances, while maintaining high success rates even when fleet size increased. By contrast, generic PPO and the heuristics failed almost completely beyond the smallest setting, and MaskPPO, although much stronger than PPO, lost reliability at larger scales. The detailed fleet-level analysis further showed that GraphPPO matched the optimization benchmark not only in reward and deliveries completed, but also through more effective charging behavior, including lower charging and waiting times than MaskPPO under shared infrastructure. In the single-truck stochastic eVRP setting, GraphPPO also remained highly competitive, indicating that its value extends beyond fleet-level congestion management to the underlying energy-feasible routing problem itself. Taken together, these results show that the combination of graph-based state--action modeling and explicit feasibility-aware action selection provides a practical and scalable approach to electric truck routing in operationally realistic settings. 

Several directions for future research follow naturally from this work. First, the current benchmark could be extended to more complex service constraints such as time windows and depot return policies. Second, the interaction between routing and charging could be modeled more deeply by incorporating reservation policies, electricity prices, larger charger heterogeneity, and explicit grid constraints, thereby advancing integrated logistics-energy decision-making. 
Finally, broader out-of-distribution evaluation and richer ablation studies would help clarify how the proposed method transfers across regions, infrastructure densities, and uncertainty regimes. 
Such extensions would further strengthen the role of learning-based methods as decision-support tools for real-world electric freight operations. 

\section*{Acknowledgements}
The study was funded by the DriVe2X research and innovation project of the European Commission, grant number 101056934. This work also used the Dutch national e-infrastructure, with support from the SURF Cooperative, under grant no. EINF-5716.


\bibliographystyle{elsarticle-num-names}

\bibliography{ref}

@inproceedings{NEURIPS2023_9bae70d3,
  author       = {Yining Ma and
                  Zhiguang Cao and
                  Yeow Meng Chee},
  editor       = {Alice Oh and
                  Tristan Naumann and
                  Amir Globerson and
                  Kate Saenko and
                  Moritz Hardt and
                  Sergey Levine},
  title        = {Learning to Search Feasible and Infeasible Regions of Routing Problems
                  with Flexible Neural k-Opt},
  booktitle    = {Advances in Neural Information Processing Systems 36: Annual Conference
                  on Neural Information Processing Systems 2023, NeurIPS 2023, New Orleans,
                  LA, USA, December 10 - 16, 2023},
  year         = {2023},
  url          = {http://papers.nips.cc/paper\_files/paper/2023/hash/9bae70d354793a95fa18751888cea07d-Abstract-Conference.html},
  bibsource    = {dblp computer science bibliography, https://dblp.org}
}

@article{BASSO2022102496,
title = {Dynamic stochastic electric vehicle routing with safe reinforcement learning},
journal = {Transportation Research Part E: Logistics and Transportation Review},
volume = {157},
pages = {102496},
year = {2022},
issn = {1366-5545},
doi = {https://doi.org/10.1016/j.tre.2021.102496},
url = {https://www.sciencedirect.com/science/article/pii/S1366554521002581},
author = {Rafael Basso and Balázs Kulcsár and Ivan Sanchez-Diaz and Xiaobo Qu}
}

@book{SuttonReinforcementIntroduction,
  title={Reinforcement learning: An introduction},
  author={Sutton, Richard S and Barto, Andrew G},
  year={2018},
  publisher={MIT press}
}

@misc{ppo,
      title={Proximal Policy Optimization Algorithms}, 
      author={John Schulman and Filip Wolski and Prafulla Dhariwal and Alec Radford and Oleg Klimov},
      year={2017},
      eprint={1707.06347},
      archivePrefix={arXiv},
      primaryClass={cs.LG},
      url={https://arxiv.org/abs/1707.06347}, 
}

@article{ALQAHTANI2022122626,
title = {Dynamic energy scheduling and routing of multiple electric vehicles using deep reinforcement learning},
journal = {Energy},
volume = {244},
pages = {122626},
year = {2022},
issn = {0360-5442},
doi = {https://doi.org/10.1016/j.energy.2021.122626},
url = {https://www.sciencedirect.com/science/article/pii/S0360544221028759},
author = {Mohammed Alqahtani and Mengqi Hu}
}

@inproceedings{kipf,
  author       = {Thomas N. Kipf and
                  Max Welling},
  title        = {Semi-Supervised Classification with Graph Convolutional Networks},
  booktitle    = {5th International Conference on Learning Representations, {ICLR} 2017,
                  Toulon, France, April 24-26, 2017, Conference Track Proceedings},
  publisher    = {OpenReview.net},
  year         = {2017},
  url          = {https://openreview.net/forum?id=SJU4ayYgl},
  bibsource    = {dblp computer science bibliography, https://dblp.org}
}

@ARTICLE{11016767,
  author={Wang, Chao and Zhang, Renyuan and Hong, Rentao and Wang, Haibo},
  journal={IEEE Transactions on Transportation Electrification}, 
  title={Attention-Enhanced Deep Reinforcement Learning for Electric Vehicle Routing Optimization}, 
  year={2025},
  volume={11},
  number={5},
  pages={11228-11242},
  doi={10.1109/TTE.2025.3574546}}

@INPROCEEDINGS{ziyan2025,
  author={Li, Ziyan and Aristov, Nikolay and Germain, Antoine and Dugundji, Elenna R.},
  booktitle={2025 IEEE High Performance Extreme Computing Conference (HPEC)}, 
  title={Multi-Stage Stochastic Programming for Heavy-Duty Electric Truck Routing Under Public Charging Congestion Uncertainty}, 
  year={2025},
  volume={},
  number={},
  pages={1-6},
  doi={10.1109/HPEC67600.2025.11196671}}

@article{5ade80e3-7757-300c-a7d3-653ed50f3347,
 ISSN = {0030364X, 15265463},
 URL = {http://www.jstor.org/stable/167074},
 author = {G. A. Croes},
 journal = {Operations Research},
 number = {6},
 pages = {791--812},
 publisher = {INFORMS},
 title = {A Method for Solving Traveling-Salesman Problems},
 urldate = {2026-04-17},
 volume = {6},
 year = {1958}
}

@inproceedings{maskppo,
  author       = {Shengyi Huang and
                  Santiago Onta{\~{n}}{\'{o}}n},
  editor       = {Roman Bart{\'{a}}k and
                  Fazel Keshtkar and
                  Michael Franklin},
  title        = {A Closer Look at Invalid Action Masking in Policy Gradient Algorithms},
  booktitle    = {Proceedings of the Thirty-Fifth International Florida Artificial Intelligence
                  Research Society Conference, {FLAIRS} 2022, Hutchinson Island, Jensen
                  Beach, Florida, USA, May 15-18, 2022},
  publisher    = {Florida Online Journals},
  year         = {2022},
  url          = {https://doi.org/10.32473/flairs.v35i.130584},
  doi          = {10.32473/FLAIRS.V35I.130584},
  bibsource    = {dblp computer science bibliography, https://dblp.org}
}

@article{sb3,
  author  = {Antonin Raffin and Ashley Hill and Adam Gleave and Anssi Kanervisto and Maximilian Ernestus and Noah Dormann},
  title   = {Stable-Baselines3: Reliable Reinforcement Learning Implementations},
  journal = {Journal of Machine Learning Research},
  year    = {2021},
  volume  = {22},
  number  = {268},
  pages   = {1-8},
  url     = {http://jmlr.org/papers/v22/20-1364.html}
}

@article{AMIRI2023109108,
title = {A robust multi-objective routing problem for heavy-duty electric trucks with uncertain energy consumption},
journal = {Computers \& Industrial Engineering},
volume = {178},
pages = {109108},
year = {2023},
issn = {0360-8352},
doi = {https://doi.org/10.1016/j.cie.2023.109108},
url = {https://www.sciencedirect.com/science/article/pii/S0360835223001328},
author = {Afsane Amiri and Hossein Zolfagharinia and Saman Hassanzadeh Amin}
}

@article{WANG2024123407,
title = {Robust routing for a mixed fleet of heavy-duty trucks with pickup and delivery under energy consumption uncertainty},
journal = {Applied Energy},
volume = {368},
pages = {123407},
year = {2024},
issn = {0306-2619},
doi = {https://doi.org/10.1016/j.apenergy.2024.123407},
url = {https://www.sciencedirect.com/science/article/pii/S0306261924007906},
author = {Ruiting Wang and Patrick Keyantuo and Teng Zeng and Jairo Sandoval and Aashrith Vishwanath and Hoseinali Borhan and Scott Moura}
}

@ARTICLE{10310266,
  author={Li, Yujing and Su, Su and Zhang, Minghao and Liu, Qiujiang and Nie, Xiaobo and Xia, Mingchao and Micu, Dan D.},
  journal={IEEE Transactions on Sustainable Energy}, 
  title={Multi-Agent Graph Reinforcement Learning Method for Electric Vehicle on-Route Charging Guidance in Coupled Transportation Electrification}, 
  year={2024},
  volume={15},
  number={2},
  pages={1180-1193},
  doi={10.1109/TSTE.2023.3330842}}

@ARTICLE{lin2025,
  author={Lin, Jian and Wang, Xintao and Niu, Rui and He, Yifan},
  journal={IEEE Transactions on Intelligent Transportation Systems}, 
  title={A Q-Learning-Based Hyper-Heuristic for Capacitated Electric Vehicle Routing Problem}, 
  year={2025},
  volume={26},
  number={10},
  pages={15746-15757},
  doi={10.1109/TITS.2025.3594393}}

@article{TANG2023121711,
title = {Energy-optimal routing for electric vehicles using deep reinforcement learning with transformer},
journal = {Applied Energy},
volume = {350},
pages = {121711},
year = {2023},
issn = {0306-2619},
doi = {https://doi.org/10.1016/j.apenergy.2023.121711},
url = {https://www.sciencedirect.com/science/article/pii/S0306261923010759},
author = {Mengcheng Tang and Weichao Zhuang and Bingbing Li and Haoji Liu and Ziyou Song and Guodong Yin}
}

@ARTICLE{10608117,
  author={Wang, Mengqin and Wei, Yanling and Huang, Xueliang and Gao, Shan},
  journal={IEEE Internet of Things Journal}, 
  title={An End-to-End Deep Reinforcement Learning Framework for Electric Vehicle Routing Problem}, 
  year={2024},
  volume={11},
  number={20},
  pages={33671-33682},
  doi={10.1109/JIOT.2024.3432911}}

@ARTICLE{9409782,
  author={Jia, Ya-Hui and Mei, Yi and Zhang, Mengjie},
  journal={IEEE Transactions on Cybernetics}, 
  title={A Bilevel Ant Colony Optimization Algorithm for Capacitated Electric Vehicle Routing Problem}, 
  year={2022},
  volume={52},
  number={10},
  pages={10855-10868},
  doi={10.1109/TCYB.2021.3069942}}

@article{SPINELLI2026105480,
title = {A stochastic electric vehicle routing problem under uncertain energy consumption},
journal = {Transportation Research Part C: Emerging Technologies},
volume = {183},
pages = {105480},
year = {2026},
issn = {0968-090X},
doi = {https://doi.org/10.1016/j.trc.2025.105480},
url = {https://www.sciencedirect.com/science/article/pii/S0968090X2500484X},
author = {Andrea Spinelli and Dario Bezzi and Ola Jabali and Francesca Maggioni}
}

@article{dong_dynamic_2023,
	title = {Dynamic electric vehicle routing problem considering mid-route recharging and new demand arrival using an improved memetic algorithm},
	volume = {58},
	issn = {2213-1388},
	url = {https://www.sciencedirect.com/science/article/pii/S2213138823003594},
	doi = {https://doi.org/10.1016/j.seta.2023.103366},
	journal = {Sustainable Energy Technologies and Assessments},
	author = {Dong, Jinting and Wang, Hongfeng and Zhang, Shuzhu},
	year = {2023},
	pages = {103366},
}

@misc{NEURIPS2018_9fb4651c,
      title={Reinforcement Learning for Solving the Vehicle Routing Problem}, 
      author={Mohammadreza Nazari and Afshin Oroojlooy and Lawrence V. Snyder and Martin Takáč},
      year={2018},
      eprint={1802.04240},
      archivePrefix={arXiv},
      primaryClass={cs.AI},
      url={https://arxiv.org/abs/1802.04240}, 
}

@ARTICLE{lin2022,
  author={Lin, Bo and Ghaddar, Bissan and Nathwani, Jatin},
  journal={IEEE Transactions on Intelligent Transportation Systems}, 
  title={Deep Reinforcement Learning for the Electric Vehicle Routing Problem With Time Windows}, 
  year={2022},
  volume={23},
  number={8},
  pages={11528-11538},
  doi={10.1109/TITS.2021.3105232}}

@article{https://doi.org/10.1155/2021/6635749,
author = {Wang, Na and Sun, Yihao and Wang, Hongfeng},
title = {An Adaptive Memetic Algorithm for Dynamic Electric Vehicle Routing Problem with Time-Varying Demands},
journal = {Mathematical Problems in Engineering},
volume = {2021},
number = {1},
pages = {6635749},
doi = {https://doi.org/10.1155/2021/6635749},
url = {https://onlinelibrary.wiley.com/doi/abs/10.1155/2021/6635749},
eprint = {https://onlinelibrary.wiley.com/doi/pdf/10.1155/2021/6635749},
year = {2021}
}

@article{euchi_hybrid_2023,
	title = {A hybrid metaheuristic algorithm to solve the electric vehicle routing problem with battery recharging stations for sustainable environmental and energy optimization},
	volume = {14},
	issn = {1868-3975},
	url = {https://doi.org/10.1007/s12667-022-00501-y},
	doi = {10.1007/s12667-022-00501-y},
	number = {1},
	journal = {Energy Systems},
	author = {Euchi, Jalel and Yassine, Adnan},
	month = feb,
	year = {2023},
	pages = {243--267},
}

@article{https://doi.org/10.1049/pel2.12555,
author = {Vani, Batchu Veena and Kishan, Dharavath and Ahmad, Md Waseem and Reddy, Ch Rama Prakasha},
title = {An efficient optimization algorithm for electric vehicle routing problem},
journal = {IET Power Electronics},
volume = {19},
number = {1},
pages = {e12555},
doi = {https://doi.org/10.1049/pel2.12555},
url = {https://ietresearch.onlinelibrary.wiley.com/doi/abs/10.1049/pel2.12555},
year = {2023}
}

@article{KUCUKOGLU2021107650,
title = {The electric vehicle routing problem and its variations: A literature review},
journal = {Computers \& Industrial Engineering},
volume = {161},
pages = {107650},
year = {2021},
issn = {0360-8352},
doi = {https://doi.org/10.1016/j.cie.2021.107650},
url = {https://www.sciencedirect.com/science/article/pii/S0360835221005544},
author = {Ilker Kucukoglu and Reginald Dewil and Dirk Cattrysse}
}

@inproceedings{Sym-NCO,
  author       = {Minsu Kim and
                  Junyoung Park and
                  Jinkyoo Park},
  editor       = {Sanmi Koyejo and
                  S. Mohamed and
                  A. Agarwal and
                  Danielle Belgrave and
                  K. Cho and
                  A. Oh},
  title        = {Sym-NCO: Leveraging Symmetricity for Neural Combinatorial Optimization},
  booktitle    = {Advances in Neural Information Processing Systems 35: Annual Conference
                  on Neural Information Processing Systems 2022, NeurIPS 2022, New Orleans,
                  LA, USA, November 28 - December 9, 2022},
  year         = {2022},
  url          = {http://papers.nips.cc/paper\_files/paper/2022/hash/0cddb777d3441326544e21b67f41bdc8-Abstract-Conference.html},
  bibsource    = {dblp computer science bibliography, https://dblp.org}
}

@inproceedings{
kool2018attention,
  author       = {Wouter Kool and
                  Herke van Hoof and
                  Max Welling},
  title        = {Attention, Learn to Solve Routing Problems!},
  booktitle    = {7th International Conference on Learning Representations, {ICLR} 2019,
                  New Orleans, LA, USA, May 6-9, 2019},
  publisher    = {OpenReview.net},
  year         = {2019},
  url          = {https://openreview.net/forum?id=ByxBFsRqYm},
  bibsource    = {dblp computer science bibliography, https://dblp.org}
}

@inproceedings{
yang2025neural,
title={Neural Combinatorial Optimization for Time Dependent Traveling Salesman Problem},
author={Ruixiao Yang and Chuchu Fan},
booktitle={The Thirty-ninth Annual Conference on Neural Information Processing Systems},
year={2025},
url={https://openreview.net/forum?id=UXTR6ZYV1x}
}

@article{orfanoudakis2026gnndt,
  title = {A Graph Neural Network Enhanced Decision Transformer for Efficient Optimization in Dynamic Smart Charging Environments},
  author = {Orfanoudakis, Stavros and Panda, Nanda Kishor and Palensky, Peter and Vergara, Pedro P.},
  journal = {Energy and AI},
  volume = {23},
  pages = {100679},
  year = {2026},
  doi = {10.1016/j.egyai.2026.100679}
}

@article{MOGALE2025111315,
title = {Modelling and optimising a multi-depot vehicle routing problem for freight distribution in a retail logistics network},
journal = {Computers \& Industrial Engineering},
volume = {207},
pages = {111315},
year = {2025},
issn = {0360-8352},
doi = {https://doi.org/10.1016/j.cie.2025.111315},
url = {https://www.sciencedirect.com/science/article/pii/S0360835225004619},
author = {D.G. Mogale and Abhijeet Ghadge and Sarat Kumar Jena}
}

@inproceedings{lombard2018modelling,
  title={Modelling the Time-dependent VRP through Open Data},
  author={Lombard, Augustin and Tamayo, Simon and Fontane, Fr{\'e}d{\'e}ric},
  booktitle={Proceedings of the International MultiConference of Engineers and Computer Scientists},
  volume={2},
  year={2018}
}

@article{brockman2016openai,
  title={Openai gym},
  author={Brockman, Greg and Cheung, Vicki and Pettersson, Ludwig and Schneider, Jonas and Schulman, John and Tang, Jie and Zaremba, Wojciech},
  journal={arXiv:1606.01540},
  year={2016}
}

@ARTICLE{10225616,
  author={Saner, Can Berk and Saha, Jaydeep and Srinivasan, Dipti},
  journal={IEEE Transactions on Intelligent Transportation Systems}, 
  title={A Charge Curve and Battery Management System Aware Optimal Charging Scheduling Framework for Electric Vehicle Fast Charging Stations With Heterogeneous Customer Mix}, 
  year={2023},
  volume={24},
  number={12},
  pages={14890-14902},
  doi={10.1109/TITS.2023.3303621}}

@book{Elsevier2013_ISBN9780124077959,
  title     = {Markov Processes for Stochastic Modeling},
  author    = {Ibe, Oliver},
  year      = {2013},
  publisher = {Elsevier},
  url       = {https://doi.org/10.1016/C2012-0-06106-6},
  language  = {English}
}

@misc{arowolo2025generalizationgraphneuralnetworks,
      title={Towards Generalization of Graph Neural Networks for AC Optimal Power Flow}, 
      author={Olayiwola Arowolo and Jochen L. Cremer},
      year={2025},
      eprint={2510.06860},
      archivePrefix={arXiv},
      primaryClass={cs.LG},
      url={https://arxiv.org/abs/2510.06860}, 
}

@article{WANG2025104932,
title = {The Heterogeneous-Fleet Electric Vehicle Routing Problem with Nonlinear Charging Functions},
journal = {Transportation Research Part C: Emerging Technologies},
volume = {170},
pages = {104932},
year = {2025},
issn = {0968-090X},
doi = {https://doi.org/10.1016/j.trc.2024.104932},
url = {https://www.sciencedirect.com/science/article/pii/S0968090X24004534},
author = {Weiquan Wang and Yossiri Adulyasak and Jean-François Cordeau and Guannan He}
}

@article{BRAGIN2024104494,
title = {Toward efficient transportation electrification of heavy-duty trucks: Joint scheduling of truck routing and charging},
journal = {Transportation Research Part C: Emerging Technologies},
volume = {160},
pages = {104494},
year = {2024},
issn = {0968-090X},
doi = {https://doi.org/10.1016/j.trc.2024.104494},
url = {https://www.sciencedirect.com/science/article/pii/S0968090X24000159},
author = {Mikhail A. Bragin and Zuzhao Ye and Nanpeng Yu}
}

@ARTICLE{ev2gym,
   title={EV2Gym: A Flexible V2G Simulator for EV Smart Charging Research and Benchmarking},
   volume={26},
   ISSN={1558-0016},
   number={2},
   journal={IEEE Transactions on Intelligent Transportation Systems},
   publisher={Institute of Electrical and Electronics Engineers (IEEE)},
   author={Orfanoudakis, Stavros and Diaz-Londono, Cesar and Emre Yılmaz, Yunus and Palensky, Peter and Vergara, Pedro P.},
   year={2025},
   month=feb, pages={2410–2421} }

@article{evgnn,
	title = {Scalable reinforcement learning for large-scale coordination of electric vehicles using graph neural networks},
	volume = {4},
	issn = {2731-3395},
	doi = {10.1038/s44172-025-00457-8},
	number = {1},
	journal = {Communications Engineering},
	author = {Orfanoudakis, Stavros and Robu, Valentin and Salazar, E. Mauricio and Palensky, Peter and Vergara, Pedro P.},
	month = jul,
	year = {2025},
	pages = {118},
}

@article{RODRIGUES2018636,
title = {Heteroscedastic Gaussian processes for uncertainty modeling in large-scale crowdsourced traffic data},
journal = {Transportation Research Part C: Emerging Technologies},
volume = {95},
pages = {636-651},
year = {2018},
issn = {0968-090X},
doi = {https://doi.org/10.1016/j.trc.2018.08.007},
url = {https://www.sciencedirect.com/science/article/pii/S0968090X18300147},
author = {Filipe Rodrigues and Francisco C. Pereira}
}

@article{lara_electric_2020,
	title = {Electric power infrastructure planning under uncertainty: stochastic dual dynamic integer programming ({SDDiP}) and parallelization scheme},
	volume = {21},
	issn = {1573-2924},
	url = {https://doi.org/10.1007/s11081-019-09471-0},
	doi = {10.1007/s11081-019-09471-0},
	number = {4},
	journal = {Optimization and Engineering},
	author = {Lara, Cristiana L. and Siirola, John D. and Grossmann, Ignacio E.},
	month = dec,
	year = {2020},
	pages = {1243--1281},
}

@techreport{osti_1615213,
  author       = {Smith, David and Ozpineci, Burak and Graves, Ronald L. and Jones, P. T. and Lustbader, Jason and Kelly, Kenneth and Walkowicz, Kevin and Birky, Alicia and Payne, Grant and Sigler, Cory and others},
  title        = {Medium- and Heavy-Duty Vehicle Electrification: An Assessment of Technology and Knowledge Gaps},
  institution  = {Oak Ridge National Laboratory (ORNL), Oak Ridge, TN (United States)},
  doi          = {10.2172/1615213},
  url          = {https://www.osti.gov/biblio/1615213},
  place        = {United States},
  year         = {2020},
  month        = {02}}

@article{ZHANG2022103152,
title = {Techno-economic comparison of electrification for heavy-duty trucks in China by 2040},
journal = {Transportation Research Part D: Transport and Environment},
volume = {102},
pages = {103152},
year = {2022},
issn = {1361-9209},
doi = {https://doi.org/10.1016/j.trd.2021.103152},
url = {https://www.sciencedirect.com/science/article/pii/S1361920921004478},
author = {Xizhao Zhang and Zhenhong Lin and Curran Crawford and Shunxi Li}
}

@article{KESKIN2021105060,
title = {A simulation-based heuristic for the electric vehicle routing problem with time windows and stochastic waiting times at recharging stations},
journal = {Computers \& Operations Research},
volume = {125},
pages = {105060},
year = {2021},
issn = {0305-0548},
doi = {https://doi.org/10.1016/j.cor.2020.105060},
url = {https://www.sciencedirect.com/science/article/pii/S0305054820301775},
author = {Merve Keskin and Bülent Çatay and Gilbert Laporte}
}

@article{cataldo_2023,
	title = {Mathematical models for the electric vehicle routing problem with time windows considering different aspects of the charging process},
	volume = {24},
	issn = {1866-1505},
	url = {https://doi.org/10.1007/s12351-023-00806-5},
	doi = {10.1007/s12351-023-00806-5},
	number = {1},
	journal = {Operational Research},
	author = {Cataldo-Díaz, Cristian and Linfati, Rodrigo and Escobar, John Willmer},
	month = dec,
	year = {2023},
	pages = {1},
}


\end{document}